\documentclass[journal]{IEEEtran}

\usepackage{mathrsfs}
\usepackage[numbers]{natbib}
\usepackage{amsmath,amsfonts}
\usepackage{amsthm}
\usepackage{array}
\usepackage{url}
\usepackage{textcomp}
\usepackage{verbatim}
\usepackage{CJK}
\usepackage{indentfirst}
\usepackage{amsthm}
\usepackage{amssymb}
\usepackage{graphicx}
\usepackage{subfigure}
\usepackage{tabularx}
\usepackage{algorithm}  
\usepackage{algpseudocode}
\usepackage{xcolor}
\usepackage{makecell}
\usepackage{ragged2e}
\usepackage[normalem]{ulem}
\usepackage{tabu}              
\usepackage{multirow}                 
\usepackage{multicol}               
\usepackage{float}                    
\usepackage{makecell}                
\usepackage{booktabs} 
\newtheorem{thm}{Theorem}
\newtheorem{remark}{Remark}

\begin{document}
\title{Evolutionary Dynamics of Variable Games in Structured Populations}

\author{Bin Pi, ~\IEEEmembership{Student Member,~IEEE,}
        Minyu Feng, ~\IEEEmembership{Senior Member,~IEEE,}
        Liang-Jian Deng, ~\IEEEmembership{Senior Member,~IEEE,}
        Xiaojie Chen,
        and Attila Szolnoki% <-this % stops a space
        
\thanks{This work was supported in part by the National Natural Science Foundation of China (NSFC) under Grant Nos. 12271083, 62273077, and 62473081, in part by the Project of the Department of Science and Technology of Sichuan Province under Grant No. 2025YFNH0001, and in part by the National Research, Development and Innovation Office (NKFIH) under Grant No. K142948.}

\thanks{Bin Pi and Xiaojie Chen are with the School of Mathematical Sciences, University of Electronic Science and Technology of China, Chengdu 611731, China (e-mail: xiaojiechen@uestc.edu.cn).

Liang-Jian Deng is with the School of Mathematical Sciences \& Multi-Hazard Early Warning Key Laboratory of Sichuan Province, University of Electronic Science and Technology of China, Chengdu 611731, China (e-mail: liangjian.deng@uestc.edu.cn).

Minyu Feng is with the College of Artificial Intelligence, Southwest University, Chongqing 400715, China.

Attila Szolnoki is with the Institute of Technical Physics and Materials Science, Centre for Energy Research, P.O. Box 49, H-1525 Budapest, Hungary.}

\thanks{Corresponding authors: Liang-Jian Deng and Xiaojie Chen.}}


\markboth{IEEE Transactions on Cybernetics}%
{Shell \MakeLowercase{\textit{et al.}}: Bare Demo of IEEEtran.cls for IEEE Journals}



\maketitle

\begin{abstract}
The game interactions among individuals in nature are often uncertain and dynamically evolving, significantly influencing the persistence of cooperation. However, it remains a formidable challenge to effectively characterize these dynamic properties in structured populations, derive theoretical conditions for cooperation, and identify the optimal game distribution for promoting cooperation. To address these issues, we propose the variable game framework in a structured population, where the game interactions between different individuals change over time. By means of the Markov chain and the pair approximation method, we derive theoretical conditions under which cooperation is favored by natural selection and when it is favored over defection under weak selection. Furthermore, we respectively formulate and solve two optimization problems to determine the optimal game distribution that most effectively fosters the evolution of cooperation by maximizing the gradient of cooperation selection and minimizing the fitness difference between defectors and cooperators. The theoretical predictions regarding both the conditions for cooperation and optimal game distribution are further validated by numerical calculations and extensive Monte Carlo simulations. Our findings offer novel insights into the mechanisms driving cooperative behavior in complex systems and provide theoretical guidance for designing optimal game environments that facilitate the evolution of cooperation.
\end{abstract}


\IEEEpeerreviewmaketitle

\vspace{-1\baselineskip}
\section{Introduction}

\IEEEPARstart{C}{ooperative} behavior is ubiquitous in nature manifesting on various scales \cite{ren2023reputation}, \cite{guan2019cooperation}, \cite{chica2023evolution}, \cite{fowler2010cooperative}, \cite{wei2025indirect} from the synergistic interactions between cells that maintain homeostasis and ensure overall health, to the sophisticated cooperative hunting strategies employed by orcas, which leverage strength and intelligence to capture larger prey, an ability that contributes to their status as apex predators. However, cooperation is often undermined by conflicts of interest among individuals, leading rational agents to favor defection strategies, ultimately resulting in the breakdown of cooperative systems. Consequently, understanding the persistence and evolution of widespread cooperation in nature has drawn considerable attention from researchers across multiple disciplines, including computer science \cite{zeng2025bursty}, \cite{tembine2009evolutionary}, mathematics \cite{wang2025optimally}, \cite{pi2024memory}, biology \cite{nowak2006five}, \cite{kasper2017genetics}, psychology \cite{henrich2021origins}, \cite{stevens2005evolving}, and so on \cite{zeng2025evolutionary}, \cite{perc2017statistical}, \cite{zhang2026effects}.

The integration of complex networks and evolutionary game theory has led to networked evolutionary games, providing a powerful framework for addressing this problem \cite{vargas2026modeling}, \cite{chica2017networked}, \cite{zeng2025complex}. In this framework, complex networks characterize the structural relationships among individuals in nature, playing a crucial role in the dissemination of information and behaviors, whereas evolutionary game theory captures the dynamics of individual strategy selection and enables the analysis of the evolutionary trends of group behavior. Since the pioneering work of Nowak and May in which they introduced the prisoner's dilemma game on a square lattice network with periodic boundaries~\cite{nowak1992evolutionary}, intensive research activity has explored the evolutionary outcomes of various game models (e.g., the snowdrift game \cite{hauert2004spatial}, \cite{li2021pool}, \cite{zhu2025adaptive}, the stag-hunt game \cite{dong2019memory}, \cite{bairagya2025coordinating}, \cite{luo2021evolutionary}, and public goods game \cite{wang2024enhancing}, \cite{pichler2017public}, \cite{han2025public}) on different types of complex networks (e.g., small-world networks \cite{chen2008smallworld}, \cite{deng2010memory}, \cite{wickramaarachchi2025uncertainty}, scale-free networks \cite{santos2005scale}, \cite{szolnoki2008scalefree}, \cite{konno2025scale}, and temporal networks \cite{holme2012temporal}, \cite{zhang2025structure}, \cite{li2020evolution}).

However, it is noteworthy that most previous studies assume a static game interaction in which individuals engage in the same game from the outset \cite{zhu2025finite}, \cite{yang2013towards}, \cite{tang2014reputation}, \cite{liu2024heterogeneously}. This assumption is idealized, as the game interactions in the real world are often uncertain and dynamically evolving. Feng {\it et al.}~\cite{feng2023evolutionary} noticed this and introduced a game transition mechanism based on a Markov process to model the continuously changing psychology of individuals in nature, demonstrating through simulations that such transitions significantly influence the evolution of cooperative behavior. Besides, Hilbe {\it et al.}~\cite{hilbe2018evolution} studied a scenario in which the availability of public resources depends on individuals' strategic choices, employing stochastic and evolutionary game theory to show that the dependence of public resources on prior interactions substantially enhances cooperative tendencies. These findings highlight the importance of studying the effects of stochastic and dynamic game interactions on the emergence and evolution of cooperation.

Although previous studies have explored games with uncertainty in individual interactions, they have primarily relied on Monte Carlo simulations or assumed that the game being played is directly influenced by individuals' strategic choices. For example, Benko {\it et al.}~\cite{benko2025evolutionary} introduced the concept of social dilemma transitions to model the evolving dilemmas faced by open data managers, finding that these transitions significantly impact cooperative behavior in open data management through extensive simulations. Additionally, Su {\it et al.}~\cite{su2019evolutionary} examined a framework in which individuals' behaviors and the game played at a given time step influence the game to be played in the subsequent step, revealing that game transitions can serve as a driving force to promote pro-social behavior in highly connected populations. However, there remains a notable gap in the theoretical analysis of game transitions, where the game interactions between individuals dynamically change over time, rather than being related to the behavior of individuals. Furthermore, it is of particular interest to investigate which conditions are most conducive to the emergence of cooperation in the context of game transitions.

To address the aforementioned problem, in this paper, we propose the variable game, where the game played between individuals is variable and dynamically evolves over time, a process that can be understood as the durations of different games played between individuals can obey various probability distributions. Our motivation for introducing this framework is to capture game variations influenced by external factors such as seasonal changes or market cycles. For instance, in the case of bulk commodities such as corn, the market dynamics shifts with the seasons. Specifically, during the summer harvest period, supply exceeds demand, placing buyers in a relatively advantageous position and driving market prices downward. Conversely, in winter, when corn supply declines, suppliers gain a competitive advantage, leading to an increase in market prices. Although this framework can be interpreted as a form of game transition, it differs fundamentally from previous studies in which game transitions are primarily driven by individual strategic choices \cite{hilbe2018evolution}, \cite{su2019evolutionary}. Specifically, the main contributions of our work are summarized as follows:

\begin{itemize}
\item We propose a novel framework of variable games in which the game interaction between different individuals evolves dynamically over time. Based on this framework, we theoretically derive the conditions under which cooperation is favored by natural selection and when it is favored over defection in a structured population.
\item We obtain the stationary game distribution of interactions between individuals based on the Markov chain, and identify the optimal game distribution that facilitates the evolution of cooperation by maximizing the gradient of cooperation selection and minimizing the fitness difference between defectors and cooperators.
\item We conduct numerical calculations and extensive Monte Carlo simulations to validate the theoretical predictions concerning both the conditions under which cooperation is favored by natural selection and when it is favored over defection, as well as the optimal game distribution. Our results demonstrate a strong consistency between theoretical analysis and simulation outcomes.
\end{itemize}

The rest of this paper is structured as follows: In Section~\ref{model}, we first introduce the evolutionary dynamics of variable games and perform theoretical analysis under weak selection to derive the conditions under which cooperation is favored by natural selection and when it is favored over defection. Subsequently, in Section~\ref{optimization}, we build two different optimization problems from distinct perspectives and theoretically determine the optimal game distribution that best promotes the evolution of cooperation. Then, we present the results of numerical calculations and Monte Carlo simulations to validate our theoretical findings in Section~\ref{simulation}. Finally, Section~\ref{conclusion} concludes the paper and provides directions for future research.

\vspace{-1\baselineskip}
\section{Evolutionary Dynamics of Variable Games}
\label{model}

In this section, we investigate the evolutionary dynamics of variable games in structured populations. We begin by outlining the game model and the strategy update rule employed by the population. Subsequently, using the pair approximation method, we derive the theoretical conditions under which cooperative behavior is favored by natural selection and when it is favored over defection.

\vspace{-1\baselineskip}
\subsection{Game Model and Strategy Update}

We study the behavioral evolution in a population in which individuals adopt cooperative or defective strategies. The population structure is modeled as a regular network in which each individual is connected to $k$ other neighbors \cite{su2019evolutionary}, \cite{han2024selection}, \cite{sakamoto2024pink}. Each individual occupies a node in the network, and the edges represent game interactions, biological reproduction, or behavioral imitation processes. At each time step, each individual interacts with its neighbors in a game. It is worth noting that the game played between an individual and a neighbor is not fixed; instead, the duration of each game interaction is governed by a random variable with an arbitrary distribution. Once the allotted time for a game elapses, it switches to another game state. This means that there exists a stationary game distribution to determine the game interaction between individuals. We denote the probability that an individual plays game $G_{i} \in \left\{ G_1, G_2, \cdots, G_n \right\}$ with a neighbor as $\pi_i$, where $\sum_{i=1}^n{\pi _i}=1$. In other words, an individual can engage in different games with different neighbors.

In this paper, we primarily focus on the two-player two-strategy game as the paradigmatic model, which is widely studied in previous research \cite{zeng2025complex}, \cite{allen2017evolutionary}. When the game played between individuals is $G_i$, interactions between different strategy pairs result in distinct payoffs: mutual cooperation (defection) yields a payoff of $R_i (P_i)$ to both individuals, while cooperation-defection interactions bring a payoff of $S_i$ to the cooperator and $T_i$ to the defector. By introducing the concept of dilemma strength \cite{wang2015universal}, \cite{fahimur2024investigating}, we can denote the payoff matrix of the game $G_i$ played between individuals as

\begin{small}
\begin{equation}
M_i=\left( \begin{matrix}
	R_i&		P_i-Dr_i\\
	R_i+Dg_i&		P_i\\
\end{matrix} \right),
\end{equation}
\end{small}
where $Dg_i$ indicates the gamble-intending dilemma and we have $Dg_i = T_i - R_i$, whereas $Dr_i$ represents the risk-averting dilemma and we have $Dr_i = P_i - S_i$. The gamble-intending ($Dg_i$) and risk-averting ($Dr_i$) dilemmas measure the extent to which individuals tend to exploit each other and the degree to which they should avoid exploitation, respectively. However, if the values of $R_i - P_i$ are significantly larger than those of $Dg_i$ and $Dr_i$, resulting in a situation similar to the limit where $T_i \rightarrow R_i$ and $P_i \rightarrow S_i$, the payoffs for each individual become independent of their own strategy choice and are instead entirely determined by the strategies of their neighbors. To address this issue, we assign $R_i = 1$ and $P_i = 0$, indicating the scaled value for the best reward and the harshest punishment, respectively. Additionally, we constrain both $Dg_i$ and $Dr_i$ within the range of $[-1, 1]$, ensuring a proportional relationship among $R_i - P_i$, $Dg_i$ and $Dr_i$. Under these conditions, the evolutionary dynamics depend on the sign and relative magnitude of $Dg_i$ and $Dr_i$. Therefore, for all games $G_{i} \in \left\{ G_1, G_2, \cdots, G_n \right\}$, the payoff matrix is given by

\begin{small}
\begin{equation}
M_i=\left( \begin{matrix}
	1&		-Dr_i\\
	1+Dg_i&		0\\
\end{matrix} \right),
\end{equation}
\end{small}
where $Dg_i,Dr_i \in [-1, 1]$. When both $Dg_i$ and $Dr_i$ are positive (negative), the game $G_i$ can be regarded as a prisoner's dilemma game (harmony game). If $Dg_i$ is positive (negative) while $Dr_i$ is negative (positive), the game $G_i$ is classified as a snowdrift game (stag hunt game). After each individual interacts with all its neighbors in the specific game, it accumulates a total payoff $F$, calculated as

\begin{small}
\begin{equation}
F = \sum_{y\in \Omega}s^T M_y s_y,
\end{equation}
\end{small}
where $s = (1, 0)^T$ and $s = (0, 1)^T$ denote cooperative and defective strategies, respectively. Here, $\Omega$ represents the neighborhood set of the focal individual, and $M_y$ indicates the payoff matrix of the game played between the focal individual and its neighbor $y$. The accumulated payoff is then mapped to the individual's fitness according to the following rule:
\begin{small}
\begin{equation}
f = 1 - \omega + \omega F\,,
\end{equation}
\end{small}
where $\omega$ indicates the intensity of selection. We particularly focus on the scenario of weak selection, i.e., $\omega \ll 1$, since the game has only a small effect on individual fitness or it is only one of many factors that influence individual fitness \cite{fu2009evolutionary}, \cite{mcavoy2021fixation}, which is widely utilized in evolutionary biology \cite{nowak2004emergence}.

Next, we consider the strategy update of individuals using the death-birth updating process~\cite{kaveh2015duality}, \cite{allen2020transient}. In particular, after each individual gets its fitness by interacting with its neighbors, one individual from the population is randomly selected to die. Subsequently, the neighbors of the dead individual occupy the empty site with a probability proportional to their fitness.

\vspace{-1\baselineskip}
\subsection{Evolutionary Dynamics of Cooperation}

In this subsection, we study the conditions under which cooperation is favored by natural selection and when it is favored over defection in structured populations with variable games based on the pair approximation method \cite{ohtsuki2006simple}, \cite{sun2023state}. We assume that the game between individuals is not deterministic, but obeys a certain distribution, and there are two strategies $A$ and $B$ available to all individuals in the population, where $A$ and $B$ can be treated as cooperative and defective behaviors of individuals, respectively.

\begin{thm}
\label{cooperation can emerge}
For a sufficiently large population size $N$, we obtain the condition under which natural selection favors strategy~$A$, namely when the fixation probability of a single $A$-individual in a population of $N-1$ $B$-individuals, denoted by $\rho_A$, satisfies $\rho_A>1/N$. This condition can be expressed as follows
\begin{small}
\begin{equation}
\label{emergence}
3k>\left( 2k^2-2k-1 \right) \sum_{i=1}^n{\pi _iDr_i}+\left( k^2-k+1 \right) \sum_{i=1}^n{\pi _iDg_i}.
\end{equation}
\end{small}
\end{thm}

\begin{remark}
The proof is given in Section~I of the Supplementary Material. This theorem establishes the condition under which natural selection favors cooperation in a sufficiently large population by comparing the fixation probability of a single cooperator with that under neutral drift. The inequality demonstrated in Eq.~(\ref{emergence}) captures the joint effect of the network structure, characterized by degree $k$, and the expected payoffs across multiple games in the variable game, weighted by the game distribution $\pi_i$. In particular, the coefficients $2k^2 - 2k - 1$ and $k^2 - k + 1$ highlight how network topology can amplify or attenuate the effects of payoff. This result suggests that for cooperation to successfully invade, the weighted average of the gamble-intending dilemma $\sum \pi_i Dg_i$ must be sufficiently low, or similarly, the weighted average of the risk-averting dilemma $\sum \pi_i Dr_i$ must be sufficiently small. Consequently, Eq.~(\ref{emergence}) offers a theoretical guideline for designing cooperative environments: by appropriately adjusting the game distribution in variable games, it is possible to steer the evolutionary dynamics in favor of cooperation.
\end{remark}

In the case of a static (non-variable) game, the condition in Eq.~(\ref{emergence}) reduces to $3k>\left( 2k^2-2k-1 \right) Dr+\left( k^2-k+1 \right) Dg$. Within the framework of the donation game, where a cooperator incurs a cost $c$ to provide a benefit $b$ for the opponent, whereas a defector bears no cost and provides no benefit, we obtain $Dg=Dr=c/(b-c)$. Substituting this into the static-game condition yields the classic rule ``$b/c > k$'', originally proposed by Ohtsuki {\it et al.}~\cite{ohtsuki2006simple}.

It is important to emphasize that natural selection favoring cooperation does not imply that cooperation is favored over defection. Therefore, we need to make a further derivation for the condition under which cooperation is favored over defection, as established in the following theorem:

\begin{thm}
\label{cooperation can prevail}
For a sufficiently large population size $N$, we derive the condition for strategy~$A$ to be favored over strategy~$B$, i.e., $\rho_A>\rho_B$, where $\rho_B$ is the fixation probability of a single $B$-individual in a population of $N-1$ $A$-individuals. The corresponding condition is given by

\begin{small}
\begin{equation}
\label{dominance}
\sum_{i=1}^n{\pi _i\left( Dr_i+Dg_i \right)}<\frac{2}{k-1}.
\end{equation}
\end{small}
\end{thm}

\begin{remark}
The proof is provided in Section~II of the Supplementary Material. This theorem offers a concise and insightful criterion to determine when cooperation is favored over defection in structured populations. In contrast to Thm.~\ref{cooperation can emerge}, which examines the condition under which natural selection favors cooperation by considering the fixation probability of a single mutant cooperator, Thm.~\ref{cooperation can prevail} addresses a different condition, i.e., cooperation is more likely to dominate the population than defection. The inequality shown in Eq.~(\ref{dominance}) establishes a direct relationship between the sum of gamble-intending and risk-averting dilemmas $Dr_i+Dg_i$, weighted by the game distribution $\pi_i$, and the network connectivity $k$. The threshold $2/(k - 1)$ reflects that higher connectivity requires more stringent conditions to achieve cooperation. This result further underscores the dual influence of network topology and game environment in variable games on evolutionary outcomes. From a practical perspective, these findings offer guidance for designing multi-game environments: by properly tuning the game distribution in variable games such that the weighted average payoff remains below a critical threshold, we can foster the prevalence of cooperation over defection.
\end{remark}

In the case of a static (non-variable) game, the condition in Eq.~(\ref{dominance}) transforms into $\left( Dr+Dg \right)<\frac{2}{k-1}$. Under the framework of the donation game, where $Dg=Dr=c/(b-c)$, the condition for cooperation to be favored over defection simplifies to ``$b/c > k$'', which is consistent with the conclusion reached by Ohtsuki {\it et al.}~\cite{ohtsuki2006simple}.

This ends our theoretical analysis of evolutionary dynamics under the proposed variable game framework. Thms.~\ref{cooperation can emerge} and \ref{cooperation can prevail} provide the theoretical conditions under which natural selection favors cooperation and when cooperation can be favored over defection, respectively, if game interactions between individuals are not fixed, but there is a game distribution. To further explore how cooperative behavior can emerge and be sustained, in Section~\ref{optimization}, we develop and solve the optimization problem from two distinct perspectives, aiming to identify the optimal game distribution that most promotes the evolution of cooperation in structured populations.

\vspace{-1\baselineskip}
\section{Optimal Game Distribution}
\label{optimization}

In real-world scenarios, the game environment in which individuals interact can profoundly shape individual behavior and consequently influence the emergence and evolution of cooperation within a population \cite{benko2025evolutionary}, \cite{emmerich2017impact}. Under the variable game framework introduced in this study, different game distributions can yield distinct evolutionary outcomes. Therefore, it is meaningful to identify the optimal game distribution that most effectively facilitates the evolution and persistence of cooperation in structured populations. To do that, we formulate and analyze the optimization problem from two complementary theoretical perspectives, including i) maximizing the gradient of cooperation selection and ii) minimizing the fitness difference between defectors and cooperators. These two approaches provide distinct, yet converging insights into how the structure of the game environment can be strategically adjusted to foster cooperative behavior in structured populations.

\vspace{-1\baselineskip}
\subsection{Maximizing the Gradient of Cooperation Selection}

Similar to the concept of replicator dynamics in the well-mixed population, the evolutionary dynamics of cooperation characterize the direction and rate of cooperative evolution in the structured population. Based on Eqs.~(14) and (23) of the Supplementary Material, we can obtain the gradient of cooperation selection $\dot{p}_A$ as a function of the percentage of cooperators $p_A$ as
\begin{small}
\begin{equation}
\label{x derivation}
\begin{aligned}
\dot{p}_A=&\omega \frac{k-2}{k\left( k-1 \right)}p_A\left( 1-p_A \right) \left[ k-\left( k^2-k-1 \right) \sum_{i=1}^n{\pi _iDr_i} \right. \\
& \left. -\sum_{i=1}^n{\pi _iDg_i}+\left( k^2-k-2 \right) p_A\sum_{i=1}^n{\pi _i\left( Dr_i-Dg_i \right)} \right].
\end{aligned}
\end{equation}
\end{small}
Consequently, cooperative behavior can be promoted by maximizing the gradient of cooperation selection when the gamble-intending and risk-averting dilemmas are sufficiently small. We stress that the term $\omega \frac{k-2}{k\left( k-1 \right)}p_A\left( 1-p_A \right)$ is independent of the game distribution and remains positive for $p_A\in(0, 1)$ and $k\ge3$. Therefore, to determine the optimal game distribution that maximizes the gradient of cooperation selection, we formulate the following optimization problem:
\begin{small}
\begin{equation}
\label{max}
\begin{aligned}
&\max  H_1\left( \Pi \right), \\
&\mathrm{s.t.} \,\,\,\,\, \dot{p}_A=\omega F_1\left( p_A,q_{A|A} \right),
\end{aligned}
\end{equation}
\end{small}
where 
\begin{small}
\begin{equation} 
\begin{aligned}
H_1\left( \Pi \right)=&-\left( k^2-k-1 \right) \sum_{i=1}^n{\pi _iDr_i}-\sum_{i=1}^n{\pi _iDg_i}\\&+\left( k^2-k-2 \right) p_A\sum_{i=1}^n{\pi _i\left( Dr_i-Dg_i \right)}.
\nonumber
\end{aligned}
\end{equation}
\end{small}

It is worth noting that the game distribution $\Pi = (\pi_1, \pi_2, \cdots, \pi_n)$ varies with the proportion of cooperators $p_A$. We derive the optimal game distribution that maximizes the gradient of cooperation selection under two different games, as summarized in the following theorem:

\begin{thm}
\label{Theorem-Max}
Consider the optimization problem involving two different games. The optimal game distribution $\Pi^*$ that maximizes the gradient of cooperation selection for any proportion of cooperators is characterized as follows:

1) When $Dr_1+Dg_2>Dg_1+Dr_2$:

(i) If $(k^2-k-1)(Dr_1-Dr_2)+(Dg_1-Dg_2)<0$, then the optimal game distribution is $\Pi^*=(1, 0)$ for any $p_A\in(0,1)$;

(ii) If $(k^2-k-1)(Dg_1-Dg_2)+(Dr_1-Dr_2)>0$, then the optimal game distribution is $\Pi^*=(0, 1)$ for any $p_A\in(0,1)$;

(iii) If $(k^2-k-1)(Dr_1-Dr_2)+(Dg_1-Dg_2)>0$ and $(k^2-k-1)(Dg_1-Dg_2)+(Dr_1-Dr_2)<0$, then the optimal game distribution is

\begin{small}
\begin{equation}
\Pi ^*=\begin{cases}
	(0, 1), p_A\in \left( 0,p_{A}^{*} \right)\\
	(1, 0), p_A\in \left[ p_{A}^{*},1 \right)\\
\end{cases},
\end{equation}
\end{small}
where $p_A^*=[( k^2-k-1 ) ( Dr_1-Dr_2 ) +( Dg_1-Dg_2 )] / [( k^2-k-2 ) ( Dr_1-Dg_1-Dr_2+Dg_2 )]$.

2) When $Dr_1+Dg_2<Dg_1+Dr_2$:

(i) If $(k^2-k-1)(Dr_1-Dr_2)+(Dg_1-Dg_2)>0$, then the optimal game distribution is $\Pi^*=(0, 1)$ for any $p_A\in(0,1)$;

(ii) If $(k^2-k-1)(Dg_1-Dg_2)+(Dr_1-Dr_2)<0$, then the optimal game distribution is $\Pi^*=(1, 0)$ for any $p_A\in(0,1)$;

(iii) If $(k^2-k-1)(Dr_1-Dr_2)+(Dg_1-Dg_2)<0$ and $(k^2-k-1)(Dg_1-Dg_2)+(Dr_1-Dr_2)>0$, then the optimal game distribution is

\begin{small}
\begin{equation}
\Pi ^*=\begin{cases}
	(1, 0), p_A\in \left( 0,p_{A}^{*} \right)\\
	(0, 1), p_A\in \left[ p_{A}^{*},1 \right)\\
\end{cases}.
\end{equation}
\end{small}
\end{thm}

\begin{remark}
The proof is given in Section~III of the Supplementary Material. This theorem provides a comprehensive characterization of the optimal game distribution that maximizes the gradient of cooperation selection for different proportions of cooperators. A key finding is the optimal game distribution governed by the sign of a specific combination of game parameters, which implies that even minor differences between $Dr_i$ and $Dg_i$ can substantially alter the optimal strategy. In the hybrid case, the presence of a critical threshold $p_A^*$ marks a transition point in the population composition at which the optimal game distribution must be adjusted to maintain the highest selective advantage for cooperation. This underscores the importance of dynamically tuning the game environment in response to the evolving state of the population.
\end{remark}

\vspace{-1.5\baselineskip}
\subsection{Minimizing the Fitness Difference Between Defectors and Cooperators}

Next, we aim to facilitate the evolution of cooperative behavior in structured populations by minimizing the fitness difference between defectors and cooperators. This optimization problem is motivated by the crucial role that fitness plays in strategy evolution, and if defectors exhibit higher fitness than cooperators, then defectors are more likely to persist and dominate the population. Therefore, it is meaningful to explore an optimal game distribution that minimizes the fitness difference between defectors and cooperators, thereby fostering cooperative behavior when gamble-intending and risk-averting dilemmas are relatively small. Before that, we need to derive the expected fitness of cooperators $\overline{f_A}$ and defectors $\overline{f_B}$, respectively. In a structured population with a regular network, the expected fitness of cooperators $\overline{f_A}$ can be expressed as
\begin{small}
\begin{equation}
\begin{aligned}
\overline{f_{A}}=&p_B\left( 1-\omega \right) +p_B\omega [ -\left( \left( k-1 \right) q_{B|A}+1 \right) \sum_{i=1}^n{\pi _iDr_i} \\
& + \left( k-1 \right) q_{A|A} ] +p_A\omega [ -\left( k-1 \right) q_{B|A}\sum_{i=1}^n{\pi _iDr_i}  \\
& + \left( \left( k-1 \right) q_{A|A}+1 \right)] +p_A\left( 1-\omega \right),
\end{aligned}
\end{equation}
\end{small}
and the expected fitness of defectors $\overline{f_B}$ is given by
\begin{small}
\begin{equation}
\begin{aligned}
\overline{f_{B}}=&p_B\left( 1-\omega \right) +p_B\omega \left[ \left( k-1 \right) q_{A|B} \left( 1 + \sum_{i=1}^n{\pi _iDg_i} \right) \right]\\
& +p_A\omega \left[ \left( \left( k-1 \right) q_{A|B}+1 \right) \left( 1 + \sum_{i=1}^n{\pi _iDg_i} \right) \right]  \\
& +p_A\left( 1-\omega \right),
\end{aligned}
\end{equation}
\end{small}
where $p_B$ is the fraction of defectors and $q_{X|Y}$ indicates the conditional probability of finding a neighbor whose strategy is $X$ given that the individual's strategy is $Y$.

Therefore, to determine the optimal game distribution that minimizes the fitness difference between defectors and cooperators, we can formulate the optimization problem as follows
\begin{small}
\begin{equation}
\label{min}
\begin{aligned}
&\min  H_2\left( \Pi \right) = \overline{f_{B}} - \overline{f_{A}}, \\
&\mathrm{s.t.} \,\,\,\,\, \dot{p}_A=\omega F_1\left( p_A,q_{A|A} \right).
\end{aligned}
\end{equation}
\end{small}

We stress that the game distribution $\Pi = (\pi_1, \pi_2, \cdots, \pi_n)$ varies with the frequency of cooperators $p_A$. In what follows, we derive the optimal game distribution that minimizes the fitness difference between defectors and cooperators under two different games, and the result is summarized in the following theorem:

\begin{thm}
\label{Theorem-Min}
Consider the optimization problem involving two different games. The optimal game distribution $\Pi^*$ that minimizes the fitness difference between defectors and cooperators for any fraction of cooperators is characterized as follows:

1) When $Dg_1+Dr_2>Dg_2+Dr_1$:

(i) If $Dr_1>Dr_2$, then the optimal game distribution is $\Pi^*=(0, 1)$ for any $p_A\in(0,1)$;

(ii) If $Dg_1<Dg_2$, then the optimal game distribution is $\Pi^*=(1, 0)$ for any $p_A\in(0,1)$;

(iii) If $Dr_1<Dr_2$ and $Dg_1>Dg_2$, then the optimal game distribution is

\begin{small}
\begin{equation}
\Pi ^*=\begin{cases}
	(1, 0), p_A\in \left( 0,p_{A}^{*} \right)\\
	(0, 1), p_A\in \left[ p_{A}^{*},1 \right)\\
\end{cases},
\end{equation}
\end{small}
where $p_A^*=\left( Dr_2-Dr_1 \right) / \left( Dg_1-Dg_2-Dr_1+Dr_2 \right)\in(0,1)$.

2) When $Dg_1+Dr_2<Dg_2+Dr_1$:

(i) If $Dr_1<Dr_2$, then the optimal game distribution is $\Pi^*=(1, 0)$ for any $p_A\in(0,1)$;

(ii) If $Dg_1>Dg_2$, then the optimal game distribution is $\Pi^*=(0, 1)$ for any $p_A\in(0,1)$;

(iii) If $Dr_1>Dr_2$ and $Dg_1<Dg_2$, then the optimal game distribution is

\begin{small}
\begin{equation}
\Pi ^*=\begin{cases}
	(0, 1), p_A\in \left( 0,p_{A}^{*} \right)\\
	(1, 0), p_A\in \left[ p_{A}^{*},1 \right)\\
\end{cases}.
\end{equation}
\end{small}
\end{thm}

\begin{remark}
The proof is provided in Section~IV of the Supplementary Material. This theorem characterizes the optimal game distribution that minimizes the fitness difference between defectors and cooperators across all population compositions. The result highlights the sensitivity of the optimal game distribution to the underlying gamble-intending and risk-averting dilemmas. Interestingly, specific scenarios also give rise to a threshold $p_A^*$, indicating a critical composition of cooperators in the population at which the system must switch its game distribution to sustain the smallest possible fitness disparity. Notably, this threshold depends solely on the payoff parameters $Dr_i$ and $Dg_i$, suggesting that effective adaptive control strategies can be devised to facilitate the emergence of cooperative behavior, even in the absence of detailed knowledge about the network structure.
\end{remark}

This concludes our theoretical analysis of the optimal game distribution under the variable game framework. Thms.~\ref{Theorem-Max} and \ref{Theorem-Min} jointly reveal two complementary design principles for steering evolutionary dynamics toward cooperative behavior. Thm.~\ref{Theorem-Max} aims to enhance the directionality of natural selection in favor of cooperation by maximizing the gradient of cooperation selection, whereas Thm.~\ref{Theorem-Min} focuses on reducing the advantage of defectors by minimizing the fitness difference between defectors and cooperators.

\vspace{-1\baselineskip}
\section{Numerical and Simulation Results}
\label{simulation}

In this section, we provide a detailed description and analysis of the simulation details as well as the numerical and simulation results to verify our theorems. In particular, in Subsection~\ref{Methods}, we first illustrate the simulation methods involved in this paper. Following this, we carry out a series of simulations to verify the theoretical conditions for cooperation under the assumption of known stationary distributions of variable games and probability distributions governing game durations, as detailed in Subsection~\ref{Game Distributions}. Then, in Subsection~\ref{Optimization Results}, we perform numerical calculations and extensive Monte Carlo simulations to validate the theoretical results derived from the optimization problems.

\vspace{-1\baselineskip}
\subsection{Methods}
\label{Methods}

\begin{algorithm}[hpt]
\caption{\textbf{Simulation of Fixation Probability}}
\label{Monte Carlo Simulation}
\begin{algorithmic}[1]
\State \textbf{Initialize} the total number of simulations $num$, the current simulation count $run = 0$, the counter for successful fixations of cooperators $num_C$ (defectors $num_D$), and the regular network structure $\mathcal{G}$, where each individual is connected to $k$ neighbors. All individuals are initially assigned to defect $D$ (cooperate $C$).
\While {$run < num$}
    \State Randomly select one individual $i$ from the population \indent \hspace{-0.3cm} and set $s_i \leftarrow C$ ($s_i \leftarrow D$) as the invading strategy.
    \While {the population has not reached an absorbing \indent \hspace{-0.3cm} state}
        \State Randomly select one individual $j$ to be replaced.
        \State One of $j$'s neighbors is selected to reproduce and \indent \indent \hspace{-0.45cm} fill the vacancy, with a probability proportional to \indent \indent \hspace{-0.45cm} their fitness.
        \If {All individuals are cooperators (defectors)}
            \State $num_{C} (num_{D})\leftarrow num_{C} (num_{D}) + 1$.
            \State $run\leftarrow run + 1$.
            \State Break.
        \EndIf
        \If {All individuals are defectors (cooperators)}
            \State $run\leftarrow run + 1$.
            \State Break.
        \EndIf
    \EndWhile
\EndWhile
\State \textbf{Compute} the fixation probability of cooperation (defection): $\rho_C (\rho_D)\leftarrow \frac{num_C (num_D)}{num}$.
\end{algorithmic}
\end{algorithm}

To validate our theoretical findings, we consider the case where the transition occurs between two different games, denoted by $G_1$ and $G_2$. Concretely, the duration of each game interaction is not fixed, but follows an arbitrary probability distribution. The game played between individuals changes once the time for the current game has elapsed. The game interaction between individuals $i$ and $j$ can thus be modeled as the Markov chain $\{X_{ij}(t), t\geq0\}$ with the state space $E = \{G_1, G_2\}$. We assume that the durations for which individuals engage in games $G_1$ and $G_2$ are governed by arbitrary distributions $g_1(t)$ and $g_2(t)$, respectively, and that the game interaction between individuals transitions to another game immediately upon the conclusion of the current one. We can get the stationary distribution of individuals playing different games based on the Markov chain theory. Specifically, the probability that the game played between individuals is $G_1$ can be expressed as

\begin{small}
\begin{equation}
\pi _1=\frac{E\left( T_{G_1} \right)}{E\left( T_{G_1} \right) +E\left( T_{G_1} \right)}=\frac{\int_0^{+\infty}{tg_1\left( t \right)}dt}{\int_0^{+\infty}{t\left[ g_1\left( t \right) +g_2\left( t \right) \right]}dt},
\end{equation}
\end{small}
where $E\left( T_{G_i} \right)$ represents the expected time of game interaction $G_i$.

Similarly, the probability that the game played between individuals is $G_2$ is given by

\begin{small}
\begin{equation}
\pi _2=\frac{E\left( T_{G_2} \right)}{E\left( T_{G_1} \right) +E\left( T_{G_1} \right)}=\frac{\int_0^{+\infty}{tg_2\left( t \right)}dt}{\int_0^{+\infty}{t\left[ g_1\left( t \right) +g_2\left( t \right) \right]}dt}.
\end{equation}
\end{small}
By substituting the derived values of $\pi_1$ and $\pi_2$ into Eqs.~(\ref{emergence}) and (\ref{dominance}), we can obtain the theoretical conditions under which natural selection favors cooperation and cooperation can be favored over defection when the distribution governing game duration is known.

The population structure is modeled as a square lattice with periodic boundaries using von~Neumann neighborhoods (where each individual has 4 neighbors, i.e., $k = 4$) or Moore neighborhoods (where each individual has 8 neighbors, i.e., $k = 8$). To simulate the fixation probability $\rho_C$  ($\rho_D$) of cooperation (defection), we initialize the population with only defectors (cooperators) and randomly select one individual to adopt the opposite strategy as an invader. The simulation proceeds until the population reaches one of the two absorbing states, i.e., all individuals in the population are either cooperators or defectors. The fixation probability $\rho_C$  ($\rho_D$) is then calculated as the proportion of simulations in which a single cooperator (defector) successfully takes over the entire population, based on $5\times10^5$ independent Monte Carlo simulations. The detailed simulation procedure for computing the fixation probability is presented in Algorithm~\ref{Monte Carlo Simulation}.

\vspace{-1\baselineskip}
\subsection{Evolution of Cooperation with Variable Games}
\label{Game Distributions}

\begin{center}
\begin{figure}[htbp]
\centering
\includegraphics[scale = 0.37]{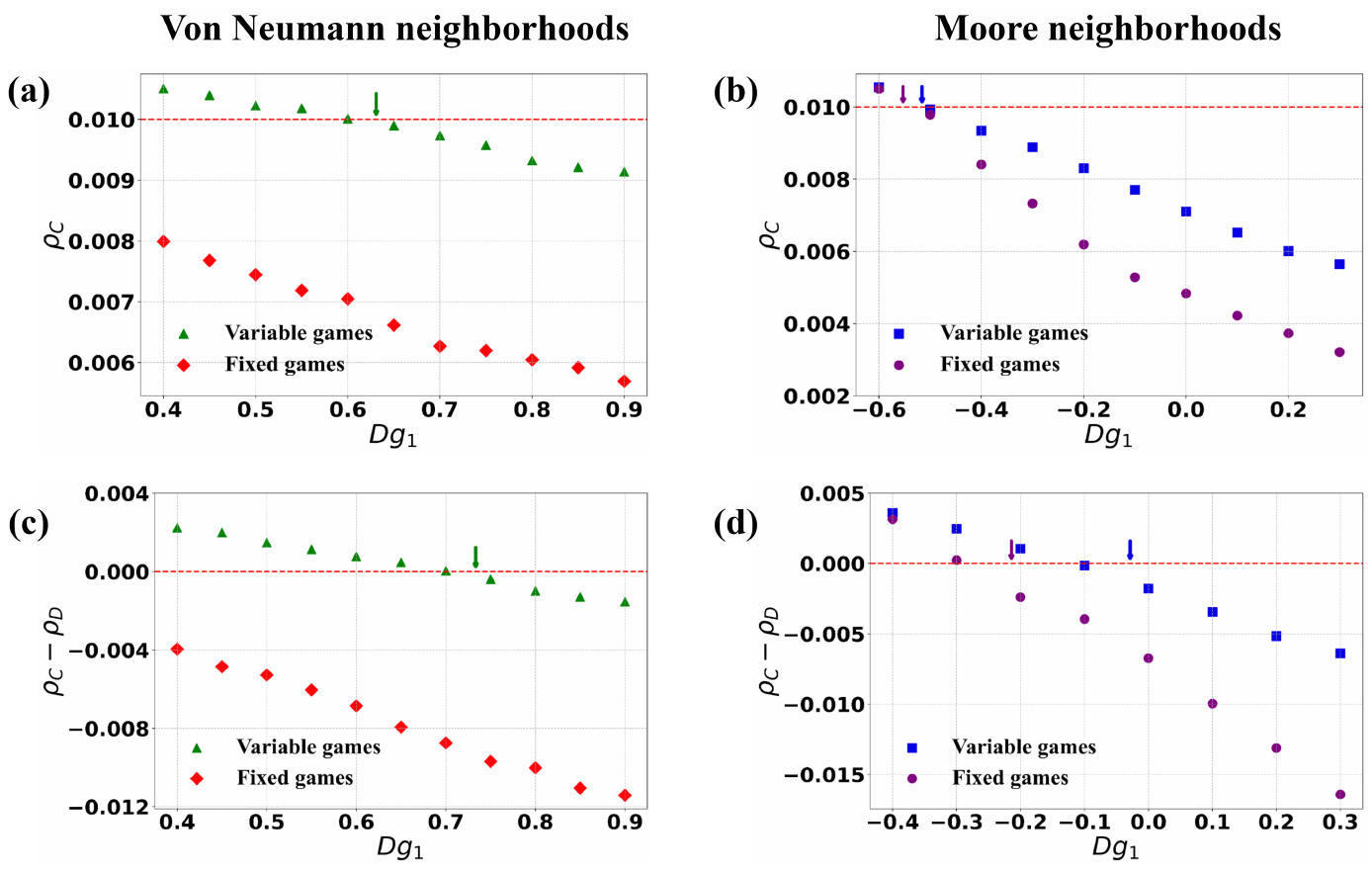}
\caption{\textbf{Fixation probability of evolutionary dynamics with known game distributions.} This figure shows $\rho_C$ and $\rho_C - \rho_D$ as functions of $Dg_1$ under different regular networks and game distributions. The arrows in the first and second rows indicate the theoretical conditions under which natural selection favors cooperation ($\rho_C>1/N$) and cooperation is favored over defection ($\rho_C>\rho_D$) in different scenarios, respectively. The horizontal red dashed line represents the neutral drift, corresponding to $\rho_C=1/N$ for panels~(a) and (b) and $\rho_C=\rho_D$ for panels~(c) and (d).}
\label{fixed distribution}
\end{figure}
\end{center}

In this subsection, we consider the scenario in which the stationary distribution of transitions between two different games, $G_1$ and $G_2$, is known and denoted by $\pi_1$ and $\pi_2$, respectively. To verify the theoretical predictions regarding the evolution of cooperation under given stationary distributions demonstrated in Thms.~\ref{cooperation can emerge} and \ref{cooperation can prevail}, we conduct different simulations on square lattice networks with von~Neumann neighborhoods ($k = 4$) and Moore neighborhoods ($k = 8$), respectively, where the network size is fixed at $N = 100$. The corresponding results are presented in Fig.~\ref{fixed distribution}, which depicts $\rho_C$ and $\rho_C - \rho_D$ as functions of $Dg_1$ under different network structures and stationary distributions. The variable game scenario corresponds to the distribution $\pi_1 = 0.5, \pi_2 = 0.5$, while the fixed game scenario corresponds to $\pi_1 = 1.0, \pi_2 = 0.0$. The arrows indicate the theoretical thresholds derived from Eqs.~(\ref{emergence}) and (\ref{dominance}), representing the conditions under which natural selection favors cooperation and cooperation is favored over defection, respectively, across different regular networks and stationary distributions. The red horizontal line marks the baseline of neutral drift for reference.

\vspace{-1\baselineskip}
\vspace{-1\baselineskip}

From Fig.~\ref{fixed distribution}, we clearly observe that both the fixation probability of cooperation $\rho_C$ and the difference $\rho_C - \rho_D$ decrease with an increase in the payoff parameter $Dg_1$. This is because an increase in $Dg_1$ leads to a greater fitness for defectors, which in turn reduces the number of cooperators in the population. In addition, by comparing the simulation results with the theoretical critical values, we find that they are in relatively good agreement. Specifically, when the payoff parameter $Dg_1$ is less than the theoretical value marked by the arrow, the cooperation shown in Figs.~\ref{fixed distribution}(a) and \ref{fixed distribution}(b) can be favored by natural selection, i.e., $\rho_C > 1 /N$, and the cooperation illustrated in Figs.~\ref{fixed distribution}(c) and \ref{fixed distribution}(d) is favored over defection, i.e., $\rho_C - \rho_D > 0$. Furthermore, we find that the points labeled by green triangles in Figs.~\ref{fixed distribution}(a) and \ref{fixed distribution}(c) are always above the red diamonds in the square lattice network with periodic boundaries using von~Neumann neighborhoods. This suggests that the introduction of variable games better facilitates the emergence and maintenance of cooperation compared to the evolution of cooperation in fixed games. In the case of Moore's neighborhoods presented in Figs.~\ref{fixed distribution}(b) and \ref{fixed distribution}(d), we can obtain the same conclusion. Moreover, by comparing the theoretical thresholds of variable and fixed games marked by the arrows, we find that the blue arrow is always to the right of the purple arrow, indicating that variable games can facilitate the evolution of cooperation more than fixed games, consistent with the results observed under the von~Neumann neighborhoods.

\begin{center}
\begin{figure}[htbp]
\centering
\includegraphics[scale = 0.37]{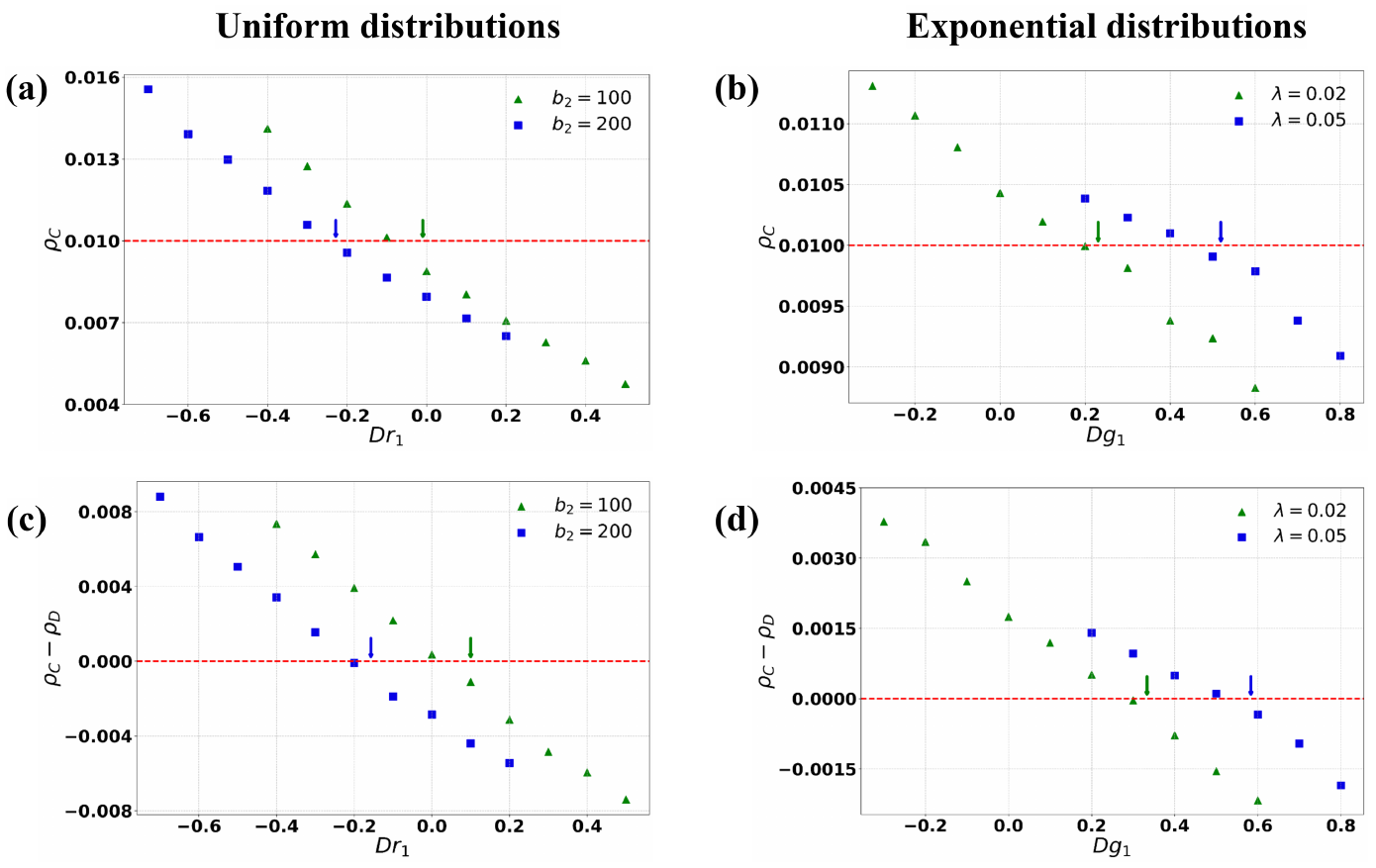}
\caption{\textbf{Fixation probability of evolutionary dynamics with the same probability distributions governing the durations of different games.} The first and second columns show the evolution of fixation probability under uniform and exponential distributions, respectively. The panels in the first row show the variation of $\rho_C$, while those in the second row discuss $\rho_C - \rho_D$. The green and blue arrows indicate the theoretical conditions for cooperation to be favored by natural selection ($\rho_C>1/N$) and cooperation to be favored over defection ($\rho_C>\rho_D$) under different transition rates, respectively. The horizontal red dashed line denotes the baseline for neutral drift, corresponding to $\rho_C=1/N$ in panels~(a) and (b), and $\rho_C=\rho_D$ in panels~(c) and (d).}
\label{same distribution}
\end{figure}
\end{center}

In what follows, we consider the scenario in which the stationary distribution of transitions between two different games is unknown; instead, only the probability distributions governing the durations of different games are specified. Fig.~\ref{same distribution} depicts the results of the fixation probability when the durations of both games obey uniform or exponential distributions. The first and second columns correspond separately to the uniform and exponential cases, while the first and second rows illustrate how $\rho_C$ and $\rho_C - \rho_D$ vary with respect to the payoff parameters, respectively.

\vspace{-1\baselineskip}
\vspace{-1\baselineskip}

The results across all four panels in Fig.~\ref{same distribution} consistently show that the simulation outcomes align well with the theoretical predictions and that cooperation will be inhibited as the payoff parameters $Dr_1$ and $Dg_1$ increase, since higher values of these parameters improve the fitness of defectors. Further comparisons under different duration distributions yield a deeper insight. In particular, when the durations of game interactions follow a uniform distribution, where the time to play $G_i$ obeys the uniform distribution of $a_i$ to $b_i$, i.e., $T_{G_i} \sim U(a_i, b_i)$, the results shown in Figs.~\ref{same distribution}(a) and \ref{same distribution}(c) are obtained on the lattice network of size $N = 100$ with Moore neighborhoods. The parameters related to the durations are set as $a_1 = a_2 = 50$, $b_1 = 150$, and those related to the payoffs are fixed at $Dg_1 = -0.2$, $Dg_2 = 0.3$, $Dr_2 = 0.5$. It can be observed that the point of $b_2 = 100$ marked by the green triangle is always above that of $b_2 = 200$ marked by the blue square, and the green arrow is shifted to the right of the blue, suggesting that lowering $b_2$, the upper bound of the uniform distribution for $G_2$, facilitates the evolution of cooperation. This phenomenon arises because $G_2$ imposes a stronger social dilemma than $G_1$, and reducing $b_2$ effectively shortens the average duration of the harsher game. On the other hand, when the game durations follow exponential distributions, where the time to play $G_1$ and $G_2$ obey the exponential distribution of $\lambda$ and $\mu$, i.e., $T_{G_1} \sim E(\lambda)$ and $T_{G_2} \sim E(\mu)$, the results displayed in Figs.~\ref{same distribution}(b) and \ref{same distribution}(d) are obtained on the lattice network of size $N = 100$ with von~Neumann neighborhoods. The transition rate from $G_2$ to $G_1$ is fixed at $\mu = 0.02$, and the payoff parameters are set as $Dr_1 = 0.5$, $Dg_2 = 0.2$, $Dr_2 = 0.3$. In this setting, increasing the transition rate $\lambda$ from $G_1$ to $G_2$ favors the survival of the cooperators. This effect can be explained in the same way as for the uniform case: by reducing the average duration spent in the more challenging game environment, the evolutionary pressure on cooperators is alleviated.

\begin{center}
\begin{figure}[htbp]
\centering
\includegraphics[scale = 0.37]{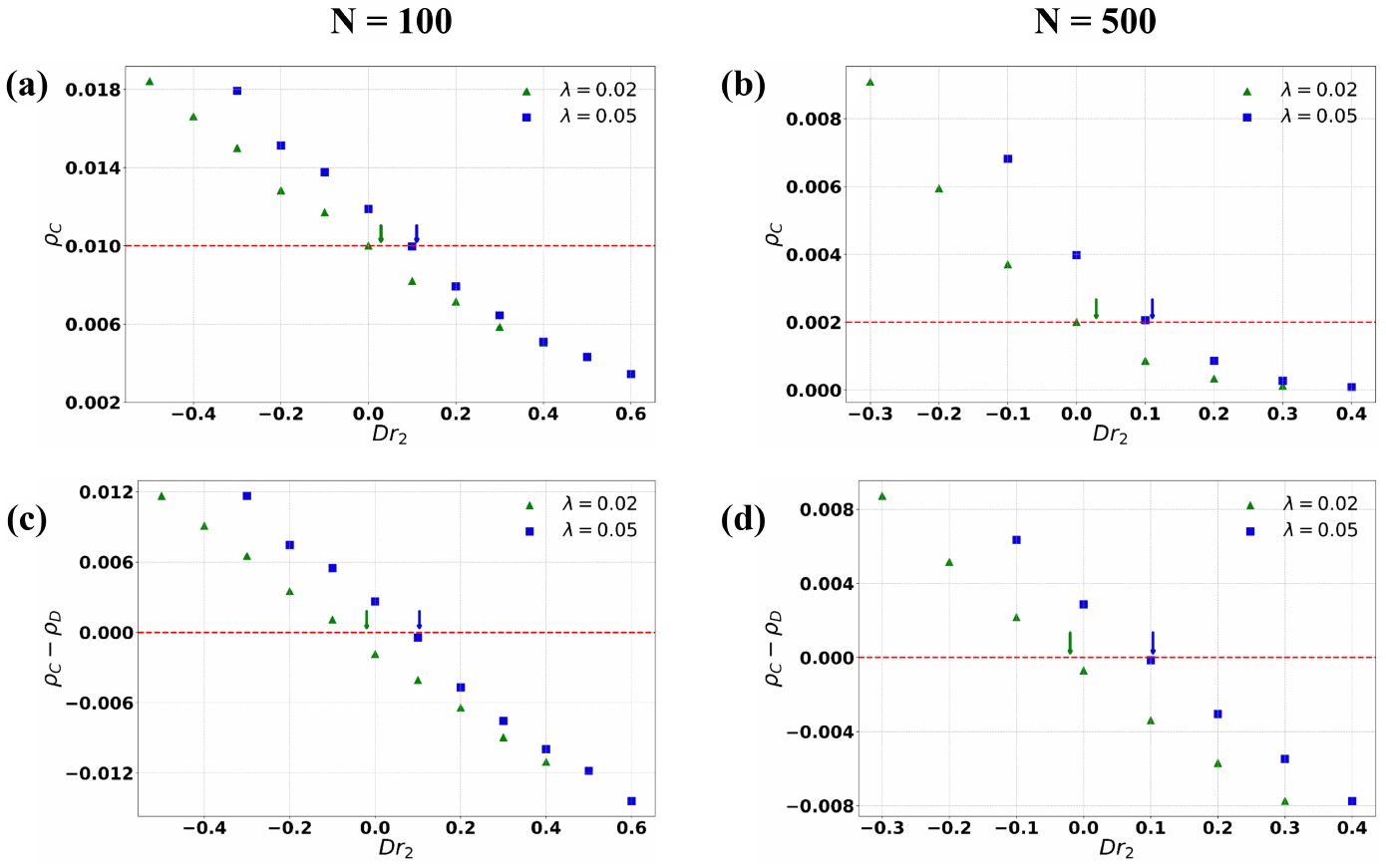}
\caption{\textbf{Fixation probability of evolutionary dynamics with arbitrary probability distributions governing the durations of different games.} The first and second rows illustrate $\rho_C$ and $\rho_C - \rho_D$ against $Dr_2$ under different exponential transition rates. The first column corresponds to a network size of $N=100$, while the second column shows the results for $N = 500$. The green and blue arrows denote the theoretical thresholds under which natural selection favors cooperation ($\rho_C>1/N$) and cooperation can be favored over defection ($\rho_C>\rho_D$) at transition rates $\lambda = 0.02$ and $\lambda = 0.05$, respectively. The horizontal red dashed line represents the neutral drift, with $\rho_C = 1/N$ in the first row and $\rho_C=\rho_D$ in the second row.}
\label{different distribution}
\end{figure}
\end{center}

Next, we explore the case where the durations for individuals to engage in games obey arbitrary, potentially distinct distributions. Unlike the previous case illustrated in Fig.~\ref{same distribution}, where both games shared the same type of distribution, here we assume that the time to play $G_1$ follows an exponential distribution with parameter $\lambda$, i.e., $T_{G_1} \sim E(\lambda)$, while the time to play game $G_2$ follows a uniform distribution over the interval $[a, b]$, i.e., $T_{G_2} \sim U(a, b)$. It is also worth noting that previous simulations were conducted on networks of size $N = 100$. In this analysis, we also examine the impact of a larger network, specifically a square lattice with $N = 500$. The parameters associated with the payoffs are set as $Dg_1 = 0.5$, $Dg_2 = 0.1$, $Dr_1 = 0.3$. The first and second rows of Fig.~\ref{different distribution} depict the evolution of $\rho_C$ and $\rho_C - \rho_D$ as a function of the payoff parameter $Dr_2$ for various values of the transition rate $\lambda$ under network sizes of $N = 100$ and $N = 500$, respectively. We emphasize that the level of neutral drift, indicated by the red horizontal dashed line in Figs.~\ref{different distribution}(a) and \ref{different distribution}(b), differs due to the distinct network sizes.

\vspace{-1\baselineskip}
\vspace{-2\baselineskip}

As shown in Fig.~\ref{different distribution}, the simulation results are in good agreement with the theoretical predictions for both network sizes, $N = 100$ and $N = 500$. It is evident that cooperation is increasingly suppressed as the payoff parameter $Dr_2$ increases. Moreover, we observe that a higher transition rate $\lambda$ promotes the evolution of cooperation. Specifically, the data point corresponding to $\lambda = 0.05$, indicated by the blue square, consistently lies above the one for $\lambda = 0.02$, marked by the green triangle. Similarly, the theoretical threshold represented by the blue arrow is always greater than that indicated by the green arrow. This observation can be explained by the fact that the social dilemma in game $G_2$ is weaker than that in $G_1$. Therefore, increasing the transition rate $\lambda$ shortens the duration of $G_1$ and effectively extends the time individuals spend playing $G_2$, which is more conducive to the persistence of cooperation. These findings are consistent with the results presented in Figs.~\ref{same distribution}(b) and \ref{same distribution}(d).

\vspace{-1\baselineskip}
\subsection{Numerical calculations and Monte Carlo Results for Optimization}
\label{Optimization Results}

In this subsection, we validate the theoretical results of the optimal game distribution derived in Section~\ref{optimization} through numerical calculations and Monte Carlo simulations.

\subsubsection{Numerical and simulation results for maximizing the gradient of cooperation selection}

\begin{center}
\begin{figure*}[htbp]
\centering
\includegraphics[scale = 0.7]{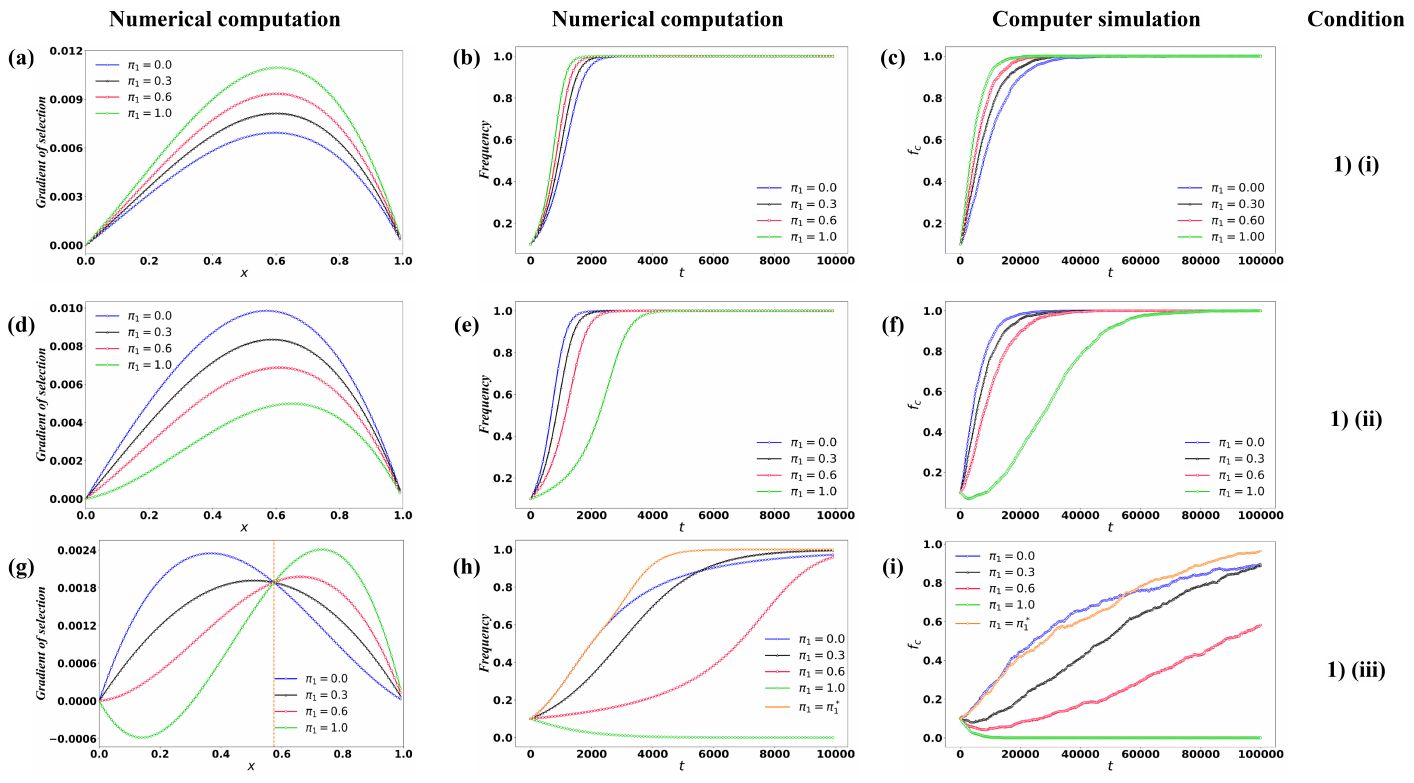}
\caption{\textbf{Numerical and simulation results for maximizing the gradient of cooperation selection.} The first column illustrates the numerical results of the gradient of cooperation selection as a function of the fraction of cooperators under different stationary game distributions. The second and third columns display the corresponding numerical and Monte Carlo simulation results, respectively, for the evolutionary trajectories of the cooperator frequency over time. Each row corresponds to a different theoretical case for maximizing the gradient of cooperation selection, as stated in items 1)(i)-(iii) of Thm.~3. The vertical orange dashed line in panel~(g) denotes the critical point $x^*$ at which $G_1(x^*) = 0$ shown in Eq.~(33) of the Supplementary Material.}
\label{gradient}
\end{figure*}
\end{center}

To verify the theoretical results concerning the maximization of the gradient of cooperation selection illustrated in Thm. \ref{Theorem-Max}, we present the numerical results of the gradient of cooperation selection as a function of the cooperator proportion, as well as the corresponding evolutionary trajectories of cooperator frequency over time, shown in the first and second columns of Fig.~\ref{gradient}, respectively. We emphasize that the numerical calculation of the temporal evolution of cooperation frequency in the second column is governed by the differential equation presented in Eq.~(\ref{x derivation}). Consequently, the corresponding evolutionary trajectories are deterministic. Our results indicate that an evolution time of $t=10000$ is sufficient to reveal the superiority of the optimal game distribution over other game distributions. Additionally, the third column of Fig.~\ref{gradient} displays Monte Carlo simulation results for the evolution of cooperator frequency over time. We note that the results in the third column are obtained from Monte Carlo simulations, which inherently introduce stochasticity into the strategy updating process at each step. Therefore, the corresponding evolutionary trajectories are non-deterministic. Since the strategy update follows a death-birth process, in which only one individual updates its strategy per time step, a sufficiently long evolutionary timescale is required to capture the asymptotic behavior of the system. Our simulations show that when the evolutionary time is set to $t = 100000$, the optimal game distribution exhibits a distinct advantage over other game distributions. The parameter configurations for rows~1-3 in Fig.~\ref{gradient} satisfy the condition $Dg_2 + Dr_1 > Dg_1 + Dr_2$, with $(k^2-k-1)(Dr_1-Dr_2)+(Dg_1-Dg_2)<0$ for row~1, $(k^2-k-1)(Dg_1-Dg_2)+(Dr_1-Dr_2)>0$ for row~2, and $(k^2-k-1)(Dr_1-Dr_2)+(Dg_1-Dg_2)>0, (k^2-k-1)(Dg_1-Dg_2)+(Dr_1-Dr_2)<0$ for row~3.

\vspace{-1\baselineskip}
\vspace{-1\baselineskip}

In Fig.~\ref{gradient}(a), we find that the gradient of selection corresponding to $\pi_1 = 1.0$ marked by green diamonds is the highest across all values of cooperator frequency and decreases monotonically as $\pi_1$ decreases. Furthermore, the result presented in Fig.~\ref{gradient}(b) shows that all scenarios eventually reach the pure cooperator state, with $\pi_1 = 1.0$ achieving this outcome the fastest. As $\pi_1$ decreases, the time required to reach full cooperation increases. The Monte Carlo simulation results in Fig.~\ref{gradient}(c) corroborate these findings. In contrast, in Fig.~\ref{gradient}(d), we find that the gradient of selection is maximized for $\pi_1 = 0.0$ marked by the blue circles and decreases as $\pi_1$ increases. Fig.~\ref{gradient}(e) further confirms that cooperation emerges most rapidly when $\pi_1 = 0.0$, compared to other situations. These observations are again supported by the simulation results illustrated in Fig.~\ref{gradient}(f). In Fig.~\ref{gradient}(g), a threshold phenomenon is observed: when the cooperator frequency $x \in (0, x^*]$, where $x^*=[( k^2-k-1 ) ( Dr_1-Dr_2 ) +( Dg_1-Dg_2 )] / [( k^2-k-2 ) ( Dr_1-Dg_1-Dr_2+Dg_2 )]$, indicated by the orange vertical dashed line, the gradient of selection is largest for $\pi_1 = 0.0$; however, when $x \in (x^*, 1)$, the gradient of selection is maximized for $\pi_1 = 1.0$. Furthermore, both the numerical and simulation results presented in Figs.~\ref{gradient}(h) and \ref{gradient}(i) demonstrate that, given a sufficiently long evolutionary time, the cooperator frequency under $\pi_1 = \pi_1^*$ always achieves the highest or reaches the full cooperation state fastest compared to the other cases, where $\pi_1^*$ denotes the value of $\pi_1$ that maximizes the gradient of selection in Fig.~\ref{gradient}(g). These results strongly validate the theoretical predictions, with both numerical and simulation outcomes displaying excellent agreement. Additionally, we verify the remaining three theoretical cases in the Supplementary Material, and the interested readers can refer to Subsection~V-A of the Supplementary Material for further details. These results also support the same conclusion: the numerical and simulation outcomes are in excellent agreement with the theoretical predictions.

\subsubsection{Numerical and simulation results for minimizing the fitness difference between defectors and cooperators}

\begin{center}
\begin{figure*}[htbp]
\centering
\includegraphics[scale = 0.7]{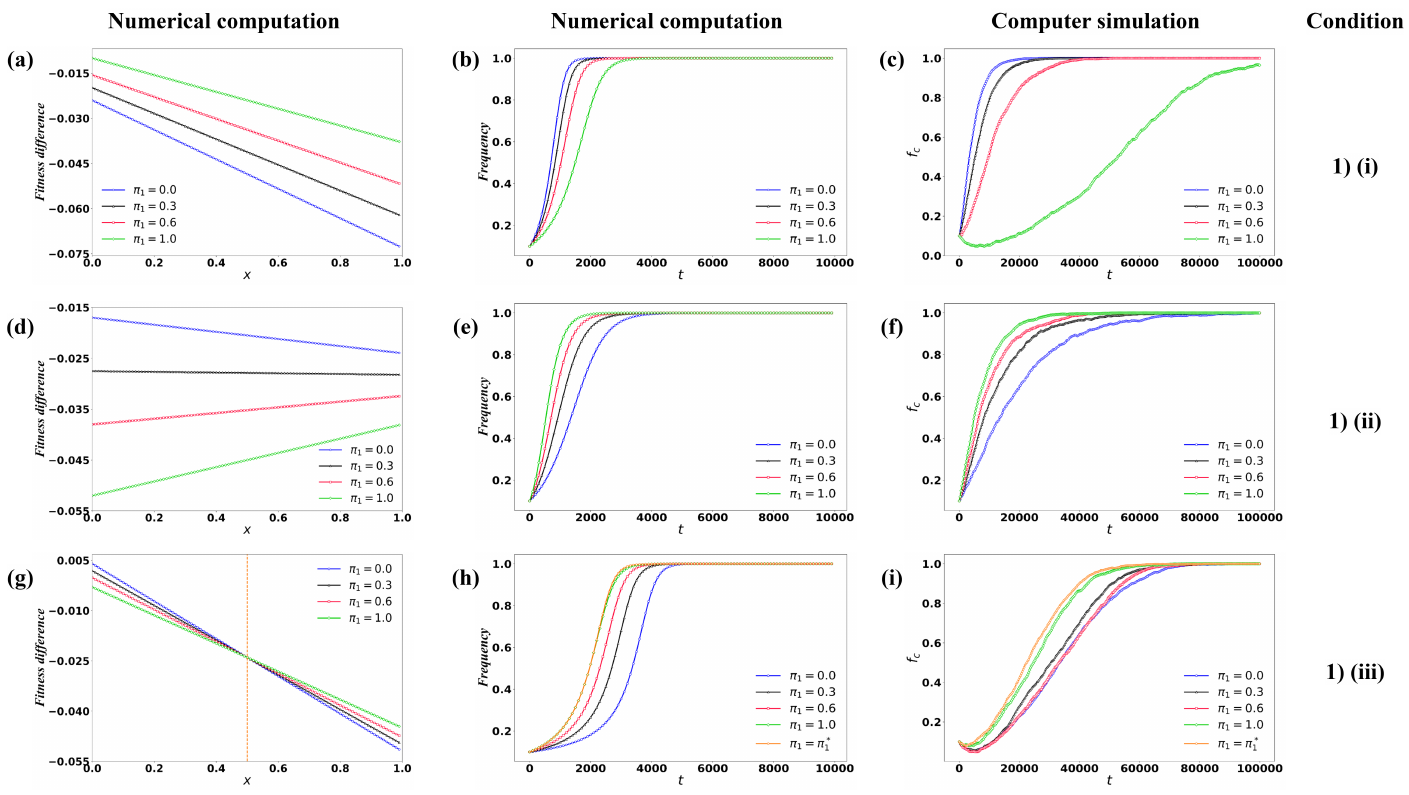}
\caption{\textbf{Numerical and simulation results for minimizing the fitness difference between defectors and cooperators.} The first column represents the numerical results of the fitness difference between defectors and cooperators as a function of the proportion of cooperators under different stationary game distributions. The second and third columns display the numerical and simulation results, respectively, for the evolutionary trajectories of the cooperator frequency over time under these distributions. Each row corresponds to a different theoretical case for minimizing the fitness difference between defectors and cooperators, as described in items~1)(i)-(iii) of Thm.~4. The orange vertical dashed line in panel~(g) indicates the critical point $x^*$ that makes $G_2(x^*) = 0$ presented in Eq.~(37) of the Supplementary Material.}
\label{fitness difference}
\end{figure*}
\end{center}

To verify the theoretical prediction of minimizing the fitness difference between defectors and cooperators shown in Thm.~\ref{Theorem-Min}, we present the numerical results of the fitness difference with respect to the proportion of cooperators, along with the evolution of the cooperator frequency over time based on Eq.~(\ref{x derivation}). These results are separately illustrated in the first and second columns of Fig.~\ref{fitness difference}. Moreover, the third column of Fig.~\ref{fitness difference} displays the Monte Carlo simulation results depicting the temporal evolution of the cooperator frequency. The evolutionary time for the numerical calculation and computer simulation of the cooperation ratio in the second and third columns is set to $t=10000$ and $t=100000$, respectively. These settings are sufficient to clearly capture the advantage of the optimal game distribution over other game distributions in promoting cooperation. The parameter settings for rows~1-3 in Fig.~\ref{fitness difference} all satisfy $Dg_1 + Dr_2 > Dg_2 + Dr_1$, with specific distinctions as follows: row 1 corresponds to $Dr_1 > Dr_2$, row 2 to $Dg_1 < Dg_2$, and row 3 to $Dr_1 < Dr_2, Dg_1 > Dg_2$.

\vspace{-1\baselineskip}
\vspace{-1\baselineskip}

In Fig.~\ref{fitness difference}(a), we observe that the fitness difference between defectors and cooperators is minimized at $\pi_1 = 0.0$ marked by blue circles and grows as $\pi_1$ increases. In addition, Fig.~\ref{fitness difference}(b) demonstrates that all scenarios eventually reach the fully cooperative state, with $\pi_1 = 0.0$ achieving it the fastest. The speed to reach full cooperation decreases as $\pi_1$ increases for the other scenarios. These findings are corroborated by the Monte Carlo simulation results displayed in Fig.~\ref{fitness difference}(c). It is worth noting that the discrepancies between the numerical calculation and computer simulation of cooperative evolution in Figs.~\ref{fitness difference}(b) and (c) are natural. This is because the numerical result is obtained from Eq.~(\ref{x derivation}), which describes a deterministic evolutionary process, whereas the computer simulation is obtained through Monte Carlo simulation, which inherently involves stochasticity in individual strategy updates, leading to indeterminate evolutionary trajectories. In Fig.~\ref{fitness difference}(d), we find that the fitness difference is minimized for $\pi_1 = 1.0$ indicated by green diamonds and increases as $\pi_1$ decreases. Analogously, Fig.~\ref{fitness difference}(e) demonstrates that the fully cooperative state is reached most rapidly when $\pi_1 = 1.0$, a conclusion also supported by the simulation results shown in Fig.~\ref{fitness difference}(f). In Fig.~\ref{fitness difference}(g), we identify a critical threshold $x^*=\left( Dr_2-Dr_1 \right) / \left( Dg_1-Dg_2-Dr_1+Dr_2 \right)$ marked by the orange vertical dashed line. When the cooperator frequency $x \in (0, x^*]$, the fitness difference is smallest for $\pi_1 = 1.0$; in contrast, for $x \in (x^*, 1)$, the smallest fitness difference occurs at $\pi_1 = 0.0$. Furthermore, the numerical and simulation results depicted in Figs.~\ref{fitness difference}(h) and \ref{fitness difference}(i) reveal that over a sufficiently long evolution period, the scenario with $\pi_1 = \pi_1^*$ achieves the highest number of cooperators or reaches full cooperation the fastest compared to the other cases, where $\pi_1^*$ corresponds to the value of $\pi_1$ that minimizes the fitness difference in Fig.~\ref{fitness difference}(g). All of these results support strong agreement between theoretical predictions and both numerical and simulation outcomes. Furthermore, the remaining three theoretical cases are validated in the Supplementary Material. Interested readers are referred to Subsection~V-B of the Supplementary Material for detailed discussions. These additional results further confirm that the numerical and simulation outcomes are in excellent agreement with the theoretical predictions.

\vspace{-1\baselineskip}
\section{Conclusions and Discussions}
\label{conclusion}

In this paper, we studied the two-player two-strategy evolutionary dynamics in structured populations under the proposed variable game framework, where the game interactions between individuals are not fixed, different from the common assumption in traditional models. In contrast, the durations of different game interactions follow a certain distribution, and we can obtain the stationary distribution of individuals engaged in various games. Based on the pair approximation method, we derived the theoretical conditions under which cooperation can be favored by natural selection and when it is favored over defection in structured populations. Notably, by applying our conclusions for the static (non-variable) game to the donation game, we recovered the classic rule ``$b/c>k$''~\cite{ohtsuki2006simple}, thereby validating the theoretical framework. To verify our theoretical results, we focused on the representative case involving transitions between two different games and performed extensive Monte Carlo simulations. The results demonstrated strong consistency between theoretical predictions and the simulation outcomes. The comparative analysis further revealed that the proposed variable game promotes cooperative behavior more effectively than the fixed game. In addition, we explored the model’s robustness by conducting simulations across different neighborhood configurations and network sizes in regular graphs. In all cases, the results remain consistent with the theoretical expectations, highlighting the robustness and generality of the proposed framework.

We also formulated two optimization problems in terms of i) maximizing the gradient of cooperation selection and ii) minimizing the fitness difference between defectors and cooperators. Through rigorous theoretical analysis, we derived the corresponding optimal game distribution that most effectively promotes the evolution of cooperation. To validate these theoretical findings, we performed numerous numerical calculations and Monte Carlo simulations for each optimization scenario. The obtained results showed that both the numerical and simulation results are in good agreement with the theoretical predictions. It is important to note that our theoretical analysis of the optimal game distribution primarily focused on the scenarios involving two distinct games. When the optimization problem extends to three or more games, obtaining an analytical solution becomes substantially more complex. In such cases, the problem can instead be addressed by numerical calculations or computational simulations. Moreover, our findings highlight that the proposed variable game mechanism exerts a substantial impact on both the fixation probability of cooperation and the conditions under which cooperation is most likely to evolve. Our work can offer valuable insights into the design of optimal game environments to foster the emergence and maintenance of cooperative behavior, particularly in contexts where modifying environmental conditions is more feasible than altering individual strategies.

Notably, the proposed variable game mechanism can be regarded as a form of game transition. However, it fundamentally differs from earlier studies, where game transitions are typically driven by individual strategic behaviors~\cite{hilbe2018evolution}, \cite{su2019evolutionary}, whereas the motivation in this paper is to model transitions influenced by external factors such as seasonal changes or market cycles. Besides, while Refs.~\cite{feng2023evolutionary} and \cite{benko2025evolutionary} investigated the effects of game transitions on the evolution of cooperation; their analysis was based on simulation experiments exclusively and did not derive the theoretical conditions under which cooperation can be favored over defection. Furthermore, we establish two optimization problems based on maximizing the gradient of cooperation selection and minimizing the fitness difference between defectors and cooperators, and theoretically derive the optimal game distribution that best promotes the evolution of cooperation. This theoretical framework represents a novel contribution, as it has not been addressed in previous works on game transitions.

Within the variable game framework proposed in this paper, several intriguing directions merit further investigation. For example, our theoretical analysis is primarily developed for regular networks, whereas population structures often exhibit other properties in reality, such as small-world effects~\cite{watts1998collective}, scale-free properties~\cite{barabasi1999emergence}, or temporal dynamics~\cite{pi2025dynamic}. Besides, our research mainly employs two-player games, whereas multi-player games are also commonly observed in practice, such as public goods games~\cite{shi2024hypergraph}. Therefore, extending the study to these more complex population structures and diverse interaction models is meaningful and remains an important avenue for future research.

%\section*{Acknowledgment}


\ifCLASSOPTIONcaptionsoff
  \newpage
\fi



%\bibliographystyle{IEEEtran}

\normalem
\bibliographystyle{ieeetr}
%\bibliography{bibtex/bib/IEEEexample}

%\vspace{-5\baselineskip}
\begin{IEEEbiography}[{\includegraphics[width=1in,height=1.25in,clip,keepaspectratio]{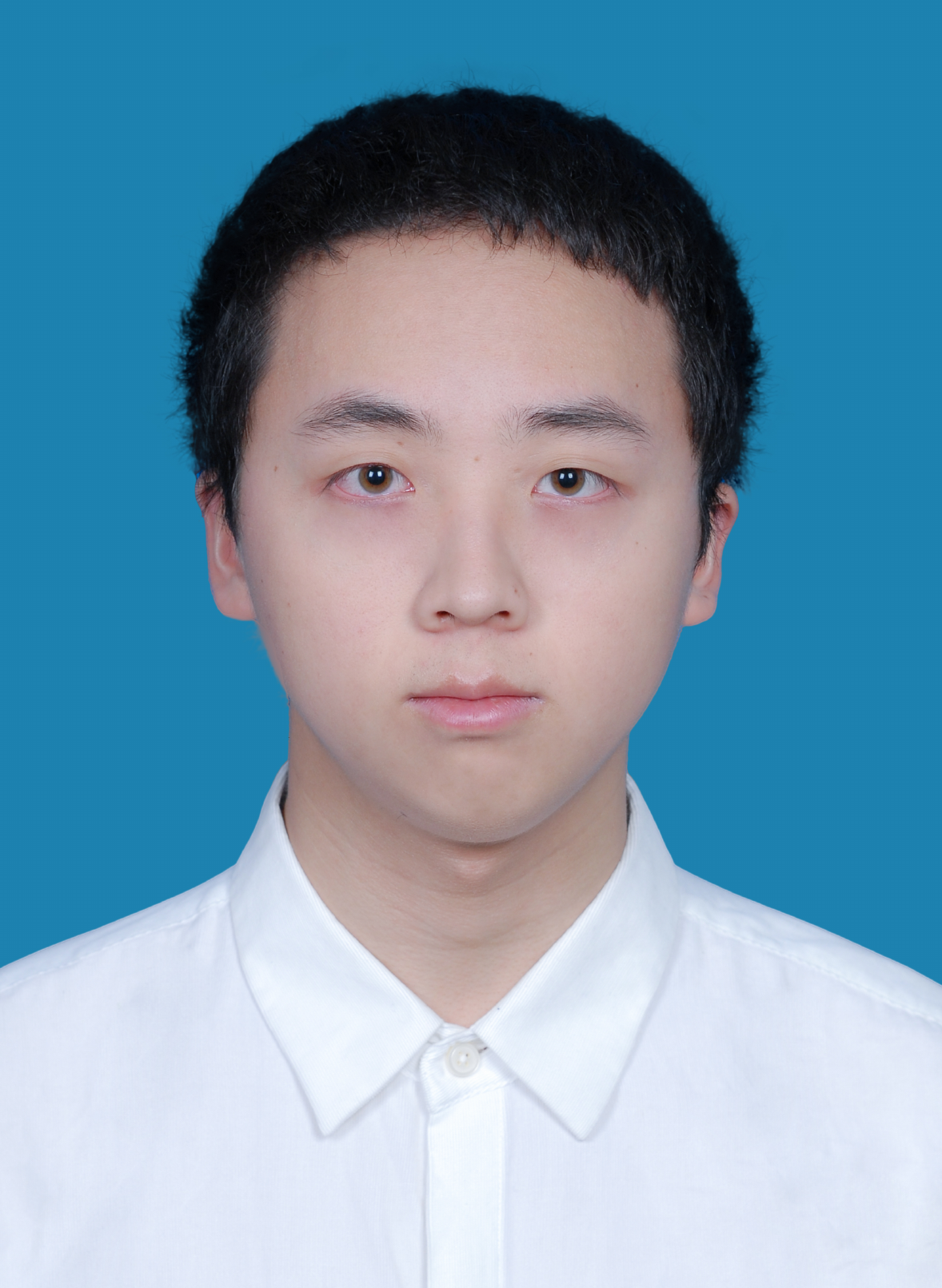}}]{Bin Pi} (Student Member, IEEE) received the B.E. degree in data science and big data technology from the College of Artificial Intelligence, Southwest University, Chongqing, China, in 2023. He is currently pursuing the M.S. degree in mathematics with the School of Mathematical Sciences, University of Electronic Science and Technology of China, Chengdu, China. His research interests include complex networks, evolutionary games, stochastic processes, and reinforcement learning.
\end{IEEEbiography}

%\vspace{-5\baselineskip}
\begin{IEEEbiography}[{\includegraphics[width=1in,height=1.25in,clip,keepaspectratio]{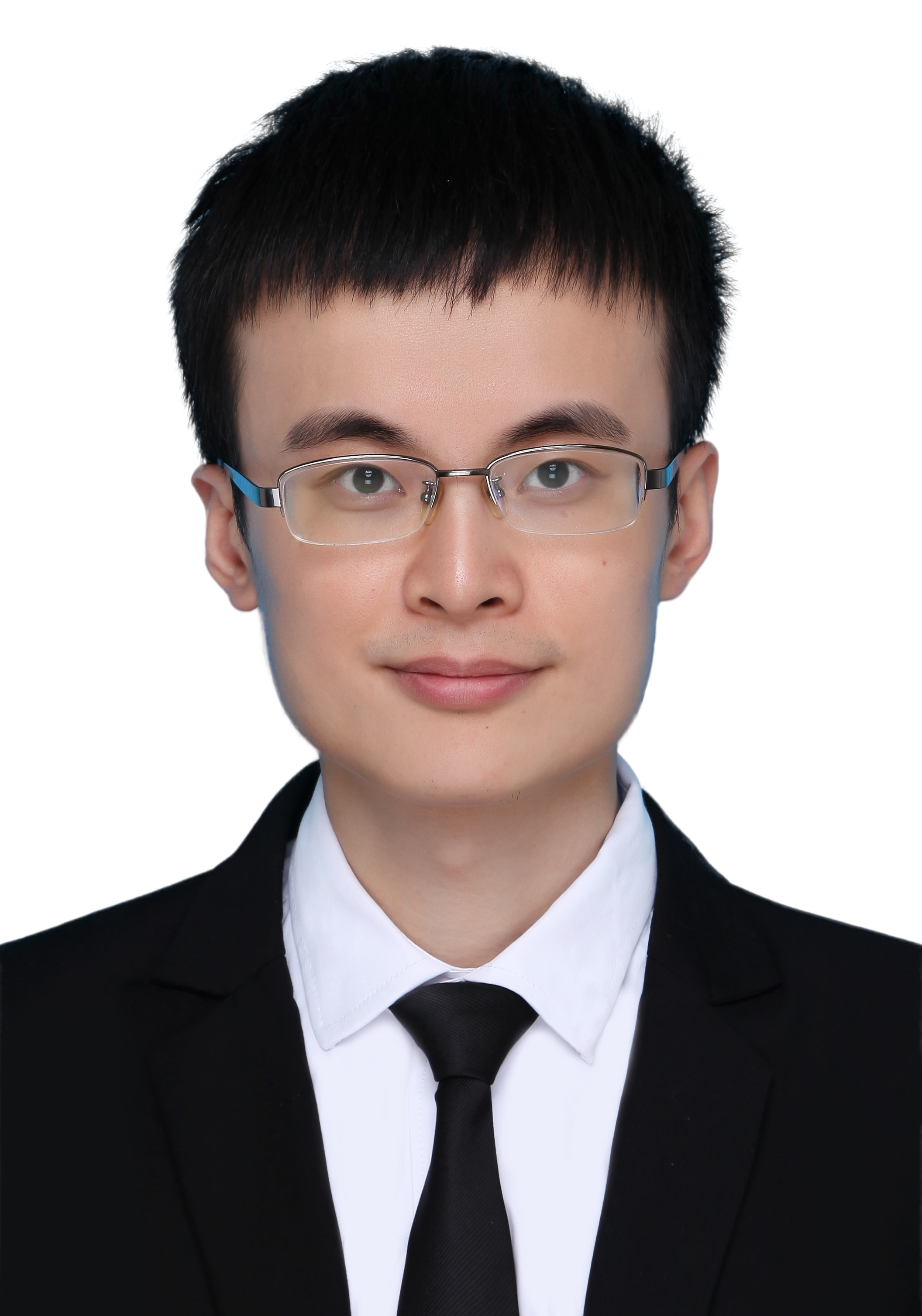}}]{Minyu Feng} (Senior Member, IEEE) received his Ph.D. degree in Computer Science from a joint program between University of Electronic Science and Technology of China, Chengdu, China, and Humboldt University of Berlin, Berlin, Germany, in 2018. Since 2019, he has been an associate professor at the College of Artificial Intelligence, Southwest University, Chongqing, China. 

Dr. Feng has published more than 80 peer-reviewed papers in authoritative journals, such as IEEE Transactions on Pattern Analysis and Machine Intelligence, IEEE Transactions on Systems, Man, and Cybernetics: Systems, IEEE Transactions on Cybernetics, etc. He is a Senior Member of China Computer Federation (CCF) and Chinese Association of Automation (CAA). Currently, he serves as a Subject Editor for Applied Mathematical Modelling, an Academic Editor for PLOS Computational Biology, an Editorial Advisory Board Member for Chaos, and an Editorial Board Member for Humanities \& Social Sciences Communications, Scientific Reports, and International Journal of Mathematics for Industry. Besides, he is a Reviewer for Mathematical Reviews of the American Mathematical Society.

Dr. Feng's research interests include Complex Systems, Evolutionary Game Theory, Computational Social Science, and Mathematical Epidemiology.
\end{IEEEbiography}

%\vspace{-5\baselineskip}
\begin{IEEEbiography}[{\includegraphics[width=1in,height=1.25in,clip,keepaspectratio]{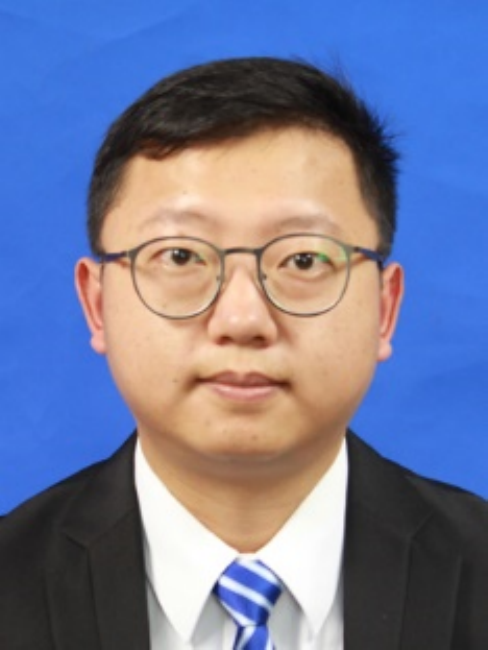}}]{Liang-Jian Deng} (Senior Member, IEEE) received the B.S. and Ph.D. degrees in applied mathematics from the School of Mathematical Sciences, University of Electronic Science and Technology of China (UESTC), Chengdu, China, in 2010 and 2016, respectively. From 2013 to 2014, he was a joint-training Ph.D. Student with Case Western Reserve University, Cleveland, OH, USA. In 2017, he was a Post-Doctoral Researcher with Hong Kong Baptist University (HKBU), Hong Kong. In addition, he has stayed at the Isaac Newton Institute for Mathematical Sciences, University of Cambridge, Cambridge, U.K., and HKBU, for short visits. He is currently a Professor with the School of Mathematical Sciences, UESTC. His research interests include use of optimization modeling, deep learning and numerical PDEs, to address several tasks in image processing and computer vision, e.g., resolution enhancement and restoration. Please visit his homepage for more info.: https://liangjiandeng.github.io/.
\end{IEEEbiography}

%\vspace{-5\baselineskip}
\begin{IEEEbiography}[{\includegraphics[width=1in,height=1.25in,clip,keepaspectratio]{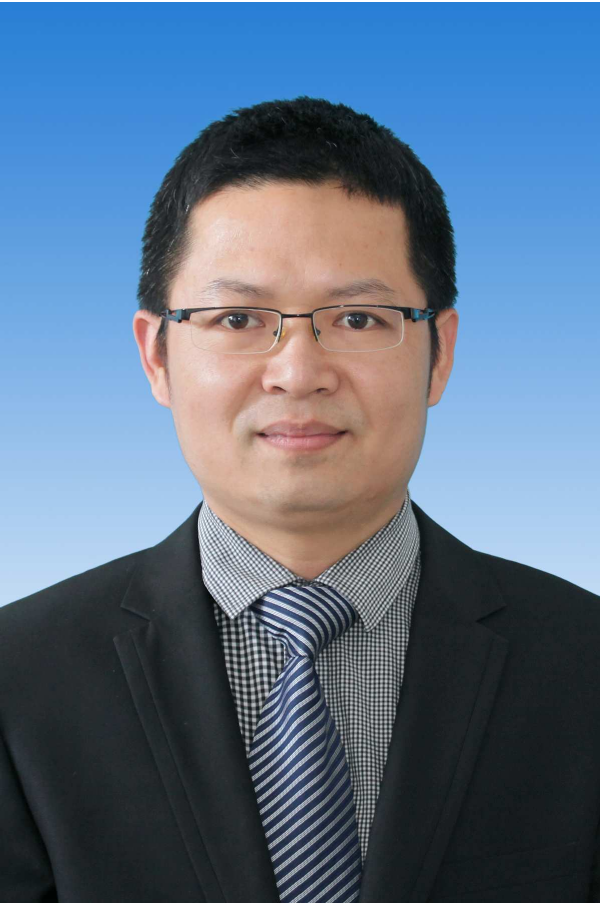}}]{Xiaojie Chen} is currently a professor at School of Mathematical Sciences in University of Electronic Science and Technology of China, China. He received the Bachelor degree in 2005 from National University of Defense Technology, China, and the PhD degree in 2011 from Peking University, China. From September 2008 to September 2009, he was a visiting scholar in University of British Columbia, Canada. From February 2011 to January 2013, he was a postdoctoral research scholar at the International Institute for Applied Systems Analysis (IIASA), Austria. From February 2013 to January 2014, he was a research scholar at IIASA, Austria. His main research interests include evolutionary dynamics, decision-making in game interactions, game-theoretical control, and collective intelligence. He has published over $100$ journal papers.
\end{IEEEbiography}

%\vspace{-5\baselineskip}
\begin{IEEEbiography}[{\includegraphics[width=1in,height=1.25in,clip,keepaspectratio]{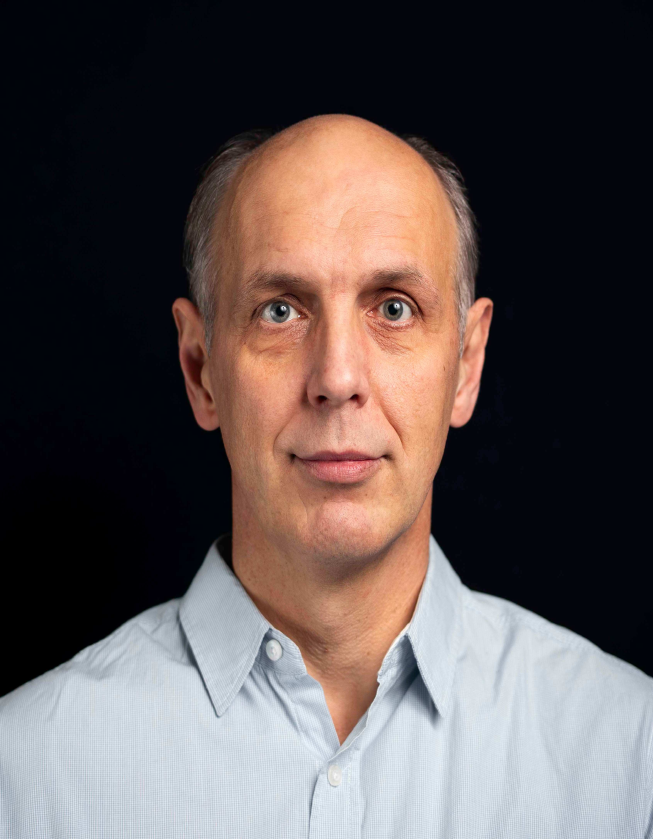}}]{Attila Szolnoki} is a research advisor with the Centre for Energy Research, Budapest, Hungary. His current research interests include evolutionary game theory, phase transitions, statistical physics and their applications. He is an Outstanding or a Distinguished Referee of several internationally recognized journals. Besides, he serves or had served as an editor for scientific journals including EPL, Physica A, Scientific Reports, Applied Mathematics and Computation, Frontiers in Physics, PLoS ONE, Entropy, or Indian Journal of Physics. He has authored around $200$ original research papers with more than 25,000 citations and an H-factor of 83. He is among top 1\% most cited physicists according to Thomson Reuters Highly Cited Researchers.
\end{IEEEbiography}


%\vfill


\end{document}

\title{Evolutionary Dynamics of Variable Games in Structured Populations (Supplementary Material)}

\author{Bin Pi, ~\IEEEmembership{Student Member,~IEEE,}
        Minyu Feng, ~\IEEEmembership{Senior Member,~IEEE,}
        Liang-Jian Deng, ~\IEEEmembership{Senior Member,~IEEE,}
        Xiaojie Chen,
        and Attila Szolnoki% <-this % stops a space
        
\thanks{This work was supported in part by the National Natural Science Foundation of China (NSFC) under Grant Nos. 12271083, 62273077, and 62473081, in part by the Project of the Department of Science and Technology of Sichuan Province under Grant No. 2025YFNH0001, and in part by the National Research, Development and Innovation Office (NKFIH) under Grant No. K142948.}

\thanks{Bin Pi and Xiaojie Chen are with the School of Mathematical Sciences, University of Electronic Science and Technology of China, Chengdu 611731, China (e-mail: xiaojiechen@uestc.edu.cn).

Liang-Jian Deng is with the School of Mathematical Sciences \& Multi-Hazard Early Warning Key Laboratory of Sichuan Province, University of Electronic Science and Technology of China, Chengdu 611731, China (e-mail: liangjian.deng@uestc.edu.cn).

Minyu Feng is with the College of Artificial Intelligence, Southwest University, Chongqing 400715, China.

Attila Szolnoki is with the Institute of Technical Physics and Materials Science, Centre for Energy Research, P.O. Box 49, H-1525 Budapest, Hungary.}

\thanks{Corresponding authors: Liang-Jian Deng and Xiaojie Chen.}}


\markboth{IEEE Transactions on Cybernetics}%
{Shell \MakeLowercase{\textit{et al.}}: Bare Demo of IEEEtran.cls for IEEE Journals}



\maketitle

\begin{abstract}
This supplementary material contains the detailed proofs of Theorems~1-4 presented in the main manuscript, along with additional figures that support and complement the main findings related to the optimization analysis.
\end{abstract}

\begin{IEEEkeywords}
Evolutionary dynamics, variable game, cooperation, structured populations.
\end{IEEEkeywords}

\IEEEpeerreviewmaketitle

\section{Proof of Theorem 1}
\label{Proof of Cooperation can Emerge}

We employ the symbols $p_A$ and $p_B$ to denote the proportions of strategies $A$ and $B$ utilized in the population. Let $p_{XY}$ and $q_{X|Y}$ respectively represent the frequency of $XY$ strategy pairs in the system and the conditional probability of finding a neighbor whose strategy is $X$ given that the individual's strategy is $Y$, where $X$ and $Y$ can be either $A$ or $B$. Then, we have the following equation holds

\begin{equation}
\label{condition}
\begin{cases}
	\begin{array}{l}
	p_A+p_B=1\\
	q_{A|X}+q_{B|X}=1\\
\end{array}\\
	p_{XY}=p_Xq_{Y|X}\\
	p_{XY}=p_{YX}\\
\end{cases},
\end{equation}
which suggests that the whole system can be described in terms of the variables $p_A$ and $q_{A|A}$, with all other variables being functions of these two variables. We emphasize that the strategies of individuals in the system evolve dynamically during the evolutionary process, with a primary focus on the frequency of strategy A utilized by individuals. Next, we derive the dynamical equation for the evolution of cooperation in the population and further identify the conditions under which cooperation dominates.

\subsection{Updating a $B$-individual}

We begin by considering the case in which a $B$-individual is selected to die, and an $A$-individual from the neighborhood succeeds in reproducing to occupy the position of the dead individual. Suppose that there are $k_A$ $A$-individuals among the $k$ neighbors of the dead $B$-individual, then the probability of this particular neighborhood configuration can be expressed as

\begin{equation}
\mathscr{A}(k_A)=\left( \begin{array}{c}
	k\\
	k_A\\
\end{array} \right) q_{A|B}^{k_A}q_{B|B}^{k_B},
\end{equation}
where $k_B$ is the number of $B$-individuals among the $k$ neighbors of the dead $B$-individual, and we have $k_B = k - k_A$.

The fitness of each $A$-individual that plays the game with the dead $B$-individual is

\begin{equation}
f_A=1-\omega +\omega \left[ \left( k-1 \right) q_{A|A}-\left( \left( k-1 \right) q_{B|A}+1 \right) \sum_{i=1}^n{\pi _iDr_i} \right],
\end{equation}
and the fitness of each $B$-individual that plays the game with the dead $B$-individual is

\begin{equation}
f_B=1-\omega +\omega \left[ \left( k-1 \right) q_{A|B}\left( 1+\sum_{i=1}^n{\pi _iDg_i} \right) \right].
\end{equation}

The probability that one of the neighboring $A$-individuals replaces the dead $B$-individual can be expressed as

\begin{equation}
P\left( A\rightarrow B \right) =\frac{k_Af_A}{k_Af_A+k_Bf_B}.
\end{equation}

Therefore, $p_A$ increases by $1/N$ with probability

\begin{equation}
\begin{aligned}
&P\left( \Delta p_A=\frac{1}{N} \right)=p_B\sum_{k_A+k_B=k}^{}{\mathscr{A}(k_A)}P\left( A\rightarrow B \right) \\
&=p_B\sum_{k_A+k_B=k}^{}{\frac{k!}{k_A!k_B!}q_{A|B}^{k_A}q_{B|B}^{k_B}}\frac{k_Af_A}{k_Af_A+k_Bf_B}.
\end{aligned}
\end{equation}

Meanwhile, the success of an $A$-individual in occupying the position of a dead $B$-individual leads to an increase in the number of AA-pair in the system by $k_A$, and thus $p_{AA}$ grows by $k_A/(kN/2)$ with probability

\begin{equation}
\begin{aligned}
&P\left( \Delta p_{AA}=\frac{2k_A}{kN} \right)=p_B\mathscr{A}(k_A)P\left( A\rightarrow B \right) \\
&=p_B\frac{k!}{k_A!k_B!}q_{A|B}^{k_A}q_{B|B}^{k_B}\frac{k_Af_A}{k_Af_A+k_Bf_B}.
\end{aligned}
\end{equation}

\subsection{Updating an $A$-individual}

Subsequently, we consider the situation in which an $A$-individual is chosen to die, and a $B$-individual in the neighborhood succeeds in reproducing to occupy the position of the dead individual. We assume that there are $k_A$ $A$-individuals and $k_B$ $B$-individuals among the $k$ neighbors of the dead $A$-individual, where $k_A + k_B = k$, then the probability of this specific neighborhood configuration is given by

\begin{equation}
\mathscr{B}(k_A)=\left( \begin{array}{c}
	k\\
	k_A\\
\end{array} \right) q_{A|A}^{k_A}q_{B|A}^{k_B}.
\end{equation}

The fitness of each $A$-individual who plays the game with the dead $A$-individual is

\begin{equation}
g_A=1-\omega +\omega \left[ \left( \left( k-1 \right) q_{A|A}+1 \right) -\left( k-1 \right) q_{B|A}\sum_{i=1}^n{\pi _iDr_i} \right],
\end{equation}
and the fitness of each $B$-individual who plays the game with the dead $A$-individual is

\begin{equation}
g_B=1-\omega +\omega \left[ \left( \left( k-1 \right) q_{A|B}+1 \right) \left( 1+\sum_{i=1}^n{\pi _iDg_i} \right) \right].
\end{equation}

The probability that one of the neighboring $B$-individuals replaces the dead $A$-individual can be expressed as

\begin{equation}
P\left( B\rightarrow A \right) =\frac{k_Bg_B}{k_Ag_A+k_Bg_B}.
\end{equation}

Therefore, $p_A$ decreases by $1/N$ with probability

\begin{equation}
\begin{aligned}
&P\left( \Delta p_A=-\frac{1}{N} \right)=p_A\sum_{k_A+k_B=k}^{}{\mathscr{B}(k_A)}P\left( B\rightarrow A \right) \\
&=p_A\sum_{k_A+k_B=k}^{}{\frac{k!}{k_A!k_B!}q_{A|A}^{k_A}q_{B|A}^{k_B}}\frac{k_Bg_B}{k_Ag_A+k_Bg_B}.
\end{aligned}
\end{equation}

Simultaneously, the success of a $B$-individual in occupying the position of a dead $A$-individual leads to a reduction in the number of AA-pair in the system by $k_A$. Consequently, $p_{AA}$ reduces by $k_A/(kN/2)$ with probability

\begin{equation}
\begin{aligned}
&P\left( \Delta p_{AA}=-\frac{2k_A}{kN} \right)=p_A\mathscr{B}(k_A)P\left( B\rightarrow A \right) \\
&=p_A\frac{k!}{k_A!k_B!}q_{A|A}^{k_A}q_{B|A}^{k_B}\frac{k_Ag_A}{k_Ag_A+k_Bg_B}.
\end{aligned}
\end{equation}

\subsection{Evolutionary Dynamics of Cooperation}

Hereby, we assume that the death of an individual and the replacement of a neighbor's strategy occur within one unit of time, $1/N$. Then, the derivative of $p_A$ can then be expressed as

\begin{small}
\begin{equation}
\label{p_A}
\begin{aligned}
&\dot{p}_A=\left[ \frac{1}{N}P\left( \Delta p_A=\frac{1}{N} \right) +\left( -\frac{1}{N} \right) P\left( \Delta p_A=-\frac{1}{N} \right) \right]N \\
&=\omega \frac{k-1}{k}p_{AB}\left( I_a-I_b\sum_{i=1}^n{\pi _iDr_i}-I_c\sum_{i=1}^n{\pi _iDg_i} \right) +O\left( \omega ^2 \right),
\end{aligned}
\end{equation}
\end{small}
where $I_a$, $I_b$, and $I_c$ are shown as follows

\begin{equation}
\label{Iabc}
\begin{cases}
	I_a=(k-1)(q_{A|A}+q_{B|B})(q_{A|A}-q_{A|B})\\
	I_b=(k-1)q_{B|A}(q_{A|A}+q_{B|B})+q_{B|B}\\
	I_c=(k-1)q_{A|B}(q_{A|A}+q_{B|B})+q_{A|A}\\
\end{cases}.
\end{equation}

Analogously, we can obtain the time derivative of the fraction of AA-pair, which is given by

\begin{equation}
\begin{aligned}
\dot{p}_{AA}&= \left[ \sum_{k_A=0}^k{\frac{2k_A}{kN}}P\left( \Delta p_{AA}=\frac{2k_A}{kN} \right) \right. \\
&\left.+\sum_{k_A=0}^k{\left( -\frac{2k_A}{kN} \right)}P\left( \Delta p_{AA}=-\frac{2k_A}{kN} \right) \right]  /\left( \frac{1}{N} \right)  \\
&=\frac{2}{k}p_{AB}\left[ 1+\left( k-1 \right) \left( q_{A|B}-q_{A|A} \right) \right] +O\left( \omega \right).  
\end{aligned}
\end{equation}

Therefore, we can derive the derivative of $q_{A|A}$

\begin{equation}
\label{q_AA}
\begin{aligned}
\dot{q}_{A|A}&=\frac{d}{dt}\left( \frac{p_{AA}}{p_A} \right) \\
&=\frac{2}{k}q_{B|A}\left[ 1+\left( k-1 \right) \left( q_{A|B}-q_{A|A} \right) \right] +O\left( \omega \right).
\end{aligned}
\end{equation}

As previously stated, the entire system can be represented solely by $p_A$ and $q_{A|A}$, i.e., Eqs.~(\ref{p_A}) and (\ref{q_AA}) can be respectively denoted as

\begin{equation}
\dot{p}_A = \omega F_1 \left(p_A, q_{A|A}\right) + O\left(\omega ^2\right),
\end{equation}
and
\begin{equation}
\dot{q}_{A|A} = F_2 \left(p_A, q_{A|A}\right) + O\left(\omega \right).
\end{equation}

When the intensity of selection is weak, i.e., $\omega \ll 1$, the local frequency $q_{A|A}$ achieves equilibrium more rapidly than the global fraction $p_A$. Consequently, the dynamical system converges swiftly to the slow manifold with $\dot{q}_{A|A} = 0$, and we get

\begin{equation}
\label{constraint}
q_{A|A} - q_{A|B} = \frac{1}{k - 1}.
\end{equation}

As we mentioned before, the whole system can be represented by $p_A$ and $q_{A|A}$. With the new constraints outlined in Eq.~(\ref{constraint}), the whole system can subsequently be characterized by a single variable $p_A$.

\subsection{Diffusion Approximation}

Next, we investigate the one-dimensional diffusion process of the random variable $p_A$. Let $\Delta p_A$ as a random variable within a very short time interval $\Delta t$. The expectation $E[\Delta p_A]$ and variance $V[\Delta p_A]$ of $\Delta p_A$ can be respectively expressed as

\begin{small}
\begin{equation}
\begin{aligned}
&E\left[ \Delta p_A \right] =\frac{1}{N}P\left( \Delta p_A=\frac{1}{N} \right) +\left( -\frac{1}{N} \right) P\left( \Delta p_A=-\frac{1}{N} \right)  \\
&=\omega \frac{k-2}{Nk}p_A\left( 1-p_A \right) \left( I_a-I_b\sum_{i=1}^n{\pi _iDr_i}-I_c\sum_{i=1}^n{\pi _iDg_i} \right) \Delta t \\
&\equiv m\left( p_A \right) \Delta t,
\end{aligned}
\end{equation}
\end{small}
and
\begin{equation}
\begin{aligned}
&V\left[ \Delta p_A \right] =E\left[ \Delta ^2p_A \right] -E^2\left[ \Delta p_A \right] \\
&=\frac{2\left( k-2 \right)}{N^2\left( k-1 \right)}p_A\left( 1-p_A \right) \Delta t\equiv v\left( p_A \right) \Delta t,
\end{aligned}
\end{equation}
where $m(p_A)$ and $v(p_A)$ are the mean and variance of the increment of $p_A$ with a very short time interval $\Delta t$. Combining Eqs.~(\ref{condition}) and (\ref{constraint}), and substituting the values of $q_{A|A}, q_{B|B}, q_{A|B}, q_{B|A}$ obtained from the solution into Eq.~(\ref{Iabc}), we yield

\begin{equation}
\label{new I}
\begin{cases}
	I_a=\frac{k}{k-1}\\
	I_b=\frac{k^2-k-1-\left( k^2-k-2 \right) p_A}{k-1}\\
	I_c=\frac{1+\left( k^2-k-2 \right) p_A}{k-1}\\
\end{cases}.
\end{equation}

The fixation probability $\phi_A(x)$ of the $A$-individual with initial ratio $p_A(t=0)=x$ follows the differential equation:

\begin{equation}
\label{fk}
m\left( x \right) \frac{d\phi _A\left( x \right)}{dx}+\frac{v\left( x \right)}{2}\frac{d^2\phi _A\left( x \right)}{dx^2}=0,
\end{equation}
and by solving Eq.~(\ref{fk}), we have
\begin{equation}
\phi _A\left( x \right) =\frac{\int_0^x{\varphi \left( y \right)}dy}{\int_0^1{\varphi \left( y \right)}dy},
\end{equation}
where $\varphi \left( y \right) =e^{-2\int_0^y{\frac{m\left( z \right)}{v\left( z \right)}dz}}$.

Therefore, under weak selection conditions, we can get

\begin{equation}
\label{fixation probability}
\begin{aligned}
\phi _A\left( x \right) =&x+\frac{\omega N}{6k}x\left( 1-x \right) \left[ \left( -2k^2+2k+1 \right) \sum_{i=1}^n{\pi _iDr_i}  \right. \\
&\left. -\left( k^2-k+1 \right) \sum_{i=1}^n{\pi _iDg_i}+3k  \right. \\
&\left. +\left( k^2-k-2 \right) x\sum_{i=1}^n{\pi _i\left( Dr_i-Dg_i \right)} \right] .
\end{aligned}
\end{equation}

\subsection{Fixation Probability}

According to Eq.~(\ref{fixation probability}), we yield the fixation probability $\rho_A$ of a single $A$-individual in a population of $N - 1$ $B$-individuals, which is given by

\begin{equation}
\label{rho_A}
\begin{aligned}
\rho_ A = &\phi _A\left( \frac{1}{N} \right) =\frac{1}{N}+\frac{\omega}{6k}\left( 1-\frac{1}{N} \right) \left[ \left( -2k^2+2k+1 \right)  \right. \\
&\left. \sum_{i=1}^n{\pi _iDr_i}-\left( k^2-k+1 \right) \sum_{i=1}^n{\pi _iDg_i}+3k \right. \\
&\left. +\frac{\left( k^2-k-2 \right)}{N}\sum_{i=1}^n{\pi _i\left( Dr_i-Dg_i \right)} \right] .
\end{aligned}
\end{equation}

Therefore, for a sufficiently large population $N$, we obtain the condition under which natural selection favors strategy A, i.e., $\rho_A>1/N$, which can be expressed as follows

\begin{equation}
3k>\left( 2k^2-2k-1 \right) \sum_{i=1}^n{\pi _iDr_i}+\left( k^2-k+1 \right) \sum_{i=1}^n{\pi _iDg_i}.
\end{equation}
$\hfill\blacksquare$

\section{Proof of Theorem 2}
\label{Proof of Cooperation can Prevail}

Based on Eq.~(\ref{fixation probability}), we can obtain the fixation probability $\rho_B$ of a single $B$-individual in a population of $N - 1$ $A$-individuals, which is given by

\begin{equation}
\label{rho_B}
\begin{aligned}
\rho_ B &= 1 - \phi _A\left( \frac{N-1}{N} \right)=\frac{1}{N}-\frac{w}{6k}\left( 1-\frac{1}{N} \right) \left[ \left( -k^2+k-1 \right) \right. \\
&\left. \sum_{i=1}^n{\pi _iDr_i}-\left( 2k^2-2k-1 \right) \sum_{i=1}^n{\pi _iDg_i}+3k \right. \\
&\left. -\frac{k^2-k-2}{N}\sum_{i=1}^n{\pi _i\left( Dr_i-Dg_i \right)} \right] .
\end{aligned}
\end{equation}

Therefore, for a sufficiently large population $N$, we get the condition under which strategy B can be favored, i.e., $\rho_B>1/N$, which can be expressed as follows

\begin{equation}
3k < \left( k^2-k+1 \right) \sum_{i=1}^n{\pi _iDr_i}+\left( 2k^2-2k-1 \right) \sum_{i=1}^n{\pi _iDg_i}.
\end{equation}

Based on Eqs.~(\ref{rho_A}) and (\ref{rho_B}), we have the ratio of the fixation probabilities

\begin{small}
\begin{equation}
\begin{aligned}
\frac{\rho _A}{\rho _B}=1+w\frac{N-1}{2}\left[ -\left( k-1 \right) \sum_{i=1}^n{\pi _iDr_i}-\left( k-1 \right) \sum_{i=1}^n{\pi _iDg_i}+2 \right] .
\end{aligned}
\end{equation}
\end{small}

Consequently, for a sufficiently large population $N$, we derive the condition for strategy A to be favored over strategy B, i.e., $\rho_A>\rho_B$, which is given by

\begin{equation}
\sum_{i=1}^n{\pi _i\left( Dr_i+Dg_i \right)}<\frac{2}{k-1}.
\end{equation}
$\hfill\blacksquare$

\section{Proof of Theorem 3}
\label{Proof of Maximizing}

According to Eq.~(8) of the main manuscript, to maximize the gradient of cooperation selection under two different games, we need to minimize $J_1(\pi_1) = \pi_1 G_1(p_A)$, subject to the constraint $\pi_1 + \pi_2 = 1$, where $G_1(p_A)$ is given by

\begin{equation}
\label{G_1(p_A)}
\begin{aligned}
G_1\left( p_A \right) =&\left( k^2-k-1 \right) \left( Dr_1-Dr_2 \right) +\left( Dg_1-Dg_2 \right) \\
&-\left( k^2-k-2 \right) \left( Dr_1-Dg_1-Dr_2+Dg_2 \right) p_A.    
\end{aligned}
\end{equation}

Since $J_1(\pi_1)$ is a linear function of $\pi_1$, its optimal value is attained at either $\pi_1 = 1$ or $\pi_1 = 0$. To determine the optimal choice, we take the derivative of $G_1(p_A)$ and yield

\begin{equation}
G_1'\left( p_A \right) =-\left( k^2-k-2 \right) \left( Dr_1-Dg_1-Dr_2+Dg_2 \right).
\end{equation}

By substituting $p_A=0$ and $p_A=1$ into Eq.~(\ref{G_1(p_A)}), we obtain

\begin{equation}
G_1\left( 0 \right) =\left( k^2-k-1 \right) \left( Dr_1-Dr_2 \right) +\left( Dg_1-Dg_2 \right),
\end{equation}
and
\begin{equation}
G_1\left( 1 \right) =\left( k^2-k-1 \right) \left( Dg_1-Dg_2 \right) +\left( Dr_1-Dr_2 \right).
\end{equation}

We first analyze the case where 1) $Dr_1+Dg_2>Dg_1+Dr_2$. In this scenario, we have $G_1'(p_A)<0$, which implies that $G_1(p_A)$ decreases as $p_A$ grows.

(i) If $(k^2-k-1)(Dr_1-Dr_2)+(Dg_1-Dg_2)<0$, then $G_1(p_A)<0$ for any $p_A\in(0,1)$ since $G_1(0)<0$ and $G_1(p_A)$ decreases monotonically as $p_A$ grows. In this case, the gradient of cooperation selection is maximized when $\pi_1 = 1$.

(ii) If $(k^2-k-1)(Dg_1-Dg_2)+(Dr_1-Dr_2)>0$, then $G_1(p_A)>0$ for any $p_A\in(0,1)$ since $G_1(1)>0$ and $G_1(p_A)$ decreases monotonically as $p_A$ increases. In this case, the gradient of cooperation selection is maximized when $\pi_1 = 0$.

(iii) If $(k^2-k-1)(Dr_1-Dr_2)+(Dg_1-Dg_2)>0$ and $(k^2-k-1)(Dg_1-Dg_2)+(Dr_1-Dr_2)<0$, then there exists a point $p_A^*=[( k^2-k-1 ) ( Dr_1-Dr_2 ) +( Dg_1-Dg_2 )] / [( k^2-k-2 ) ( Dr_1-Dg_1-Dr_2+Dg_2 )]\in(0,1)$ such that $G(p_A^*)=0$ since $G_1(0)>0$, $G_1(1)<0$, and $G_1(p_A)$ decreases monotonically as $p_A$ grows. Therefore, $G_1(p_A)>0$ always holds for $p_A\in (0, p_A^*)$, and the gradient of cooperation selection is maximized when $\pi_1 = 0$. In contrast, for $p_A\in[p_A^*, 1)$, $G_1(p_A)\leq0$ always satisfies, and the gradient of cooperation selection is maximized when $\pi_1 = 1$.

Subsequently, we investigate the case where 2) $Dr_1+Dg_2<Dg_1+Dr_2$. In this scenario, we have $G_1'(p_A)>0$, which means that $G_1(p_A)$ increases as $p_A$ grows.

(i) If $(k^2-k-1)(Dr_1-Dr_2)+(Dg_1-Dg_2)>0$, then $G_1(p_A)>0$ for any $p_A\in(0,1)$ since $G_1(0)>0$ and $G_1(p_A)$ increases monotonically as $p_A$ grows. In this case, the gradient of cooperation selection is maximized when $\pi_1 = 0$.

(ii) If $(k^2-k-1)(Dg_1-Dg_2)+(Dr_1-Dr_2)<0$, then $G_1(p_A)<0$ for any $p_A\in(0,1)$ since $G_1(1)<0$ and $G_1(p_A)$ increases monotonically as $p_A$ increases. In this case, the gradient of cooperation selection is maximized when $\pi_1 = 1$.

(iii) If $(k^2-k-1)(Dr_1-Dr_2)+(Dg_1-Dg_2)<0$ and $(k^2-k-1)(Dg_1-Dg_2)+(Dr_1-Dr_2)>0$, then there exists a point $p_A^*=[( k^2-k-1 ) ( Dr_1-Dr_2 ) +( Dg_1-Dg_2 )] / [( k^2-k-2 ) ( Dr_1-Dg_1-Dr_2+Dg_2 )]\in(0,1)$ such that $G_1(p_A^*)=0$ since $G_1(0)<0$, $G_1(1)>0$, and $G_1(p_A)$ increases monotonically as $p_A$ grows. Therefore, $G_1(p_A)<0$ always holds for $p_A\in (0, p_A^*)$, and the gradient of cooperation selection is maximized when $\pi_1 = 1$. Conversely, for $p_A\in[p_A^*, 1)$, $G_1(p_A)\geq0$ always satisfies, and thereby the gradient of cooperation selection is maximized when $\pi_1 = 0$.
$\hfill\blacksquare$

\section{Proof of Theorem 4}
\label{Proof of Minimizing}

Based on Eqs.~(\ref{condition}) and (\ref{constraint}), we can transform the optimization problem into minimizing $H_2(\Pi)=\sum_{i=1}^n{\left[ p_A\left( Dg_i-Dr_i \right) +Dr_i \right] \pi _i}$. In the case of two different games, to minimize the fitness difference between defectors and cooperators, we need to minimize $J_2(\pi_1) = \pi_1 G_2(p_A)$, subject to the constraint $\pi_1 + \pi_2 = 1$, where $G_2(p_A)$ have the following form

\begin{equation}
\label{G_2(p_A)}
G_2(p_A)=(Dg_1-Dg_2)p_A+(Dr_1-Dr_2)(1-p_A).
\end{equation}

We can get the optimal value of $\pi_1$ is either 1 or 0 since $J_2(\pi_1)$ is a linear function of $\pi_1$. To determine the optimal choice, we take the derivative of $G_2(p_A)$ and yield

\begin{equation}
G_2'\left( p_A \right) = Dg_1-Dg_2+Dr_2-Dr_1.
\end{equation}

By substituting $p_A=0$ and $p_A=1$ into Eq.~(\ref{G_2(p_A)}), we have

\begin{equation}
G_2\left( 0 \right) =Dr_1-Dr_2,
\end{equation}
and
\begin{equation}
G_2\left( 1 \right) =Dg_1-Dg_2.
\end{equation}

We first analyze the situation of 1) $Dg_1+Dr_2>Dg_2+Dr_1$, where we have $G_2'(p_A)>0$, indicating that $G_2(p_A)$ increases as $p_A$ grows.

(i) If $Dr_1>Dr_2$, then $G_2(p_A)>0$ for any $p_A\in(0,1)$ since $G_2(0)>0$ and $G_2(p_A)$ increases monotonically as $p_A$ increases. In this case, the fitness difference between defectors and cooperators is minimized when $\pi_1 = 0$.

(ii) If $Dg_1<Dg_2$, then $G_2(p_A)<0$ for any $p_A\in(0,1)$ since $G_2(1)<0$ and $G_2(p_A)$ increases monotonically as $p_A$ grows. In this case, the fitness difference between defectors and cooperators is minimized when $\pi_1 = 1$.

(iii) If $Dr_1<Dr_2$ and $Dg_1>Dg_2$, then there exists a point $p_A^*=( Dr_2-Dr_1 ) / ( Dg_1-Dg_2-Dr_1+Dr_2 )\in(0,1)$ such that $G_2(p_A^*)=0$ since $G_2(0)<0$, $G_2(1)>0$, and $G_2(p_A)$ increases monotonically as $p_A$ increases. Therefore, $G_2(p_A)<0$ always holds for $p_A \in (0, p_A^*)$, and the fitness difference between defectors and cooperators is minimized when $\pi_1 = 1$. In contrast, for $p_A \in [p_A^*, 1)$, $G_2(p_A)\geq0$ always satisfies, and thereby the fitness difference between defectors and cooperators is minimized when $\pi_1 = 0$.

Subsequently, we investigate the situation of 2) $Dg_1+Dr_2<Dg_2+Dr_1$, where we have $G_2'(p_A)<0$, meaning that $G_2(p_A)$ decreases as $p_A$ grows.

(i) If $Dr_1<Dr_2$, then $G_2(p_A)<0$ for any $p_A\in(0,1)$ since $G_2(0)<0$ and $G_2(p_A)$ decreases monotonically as $p_A$ increases. In this case, the fitness difference between defectors and cooperators is minimized when $\pi_1 = 1$.

(ii) If $Dg_1>Dg_2$, then $G_2(p_A)>0$ for any $p_A\in(0,1)$ since $G_2(1)>0$ and $G_2(p_A)$ decreases monotonically as $p_A$ grows. In this case, the fitness difference between defectors and cooperators is minimized when $\pi_1 = 0$.

(iii) If $Dr_1>Dr_2$ and $Dg_1<Dg_2$, then there exists a point $p_A^*=( Dr_2-Dr_1 ) / ( Dg_1-Dg_2-Dr_1+Dr_2 )\in(0,1)$ such that $G_2(p_A^*)=0$ since $G_2(0)>0$, $G_2(1)<0$, and $G_2(p_A)$ decreases monotonically as $p_A$ grows. Therefore, $G_2(p_A)>0$ always holds for $p_A \in (0, p_A^*)$, and the fitness difference between defectors and cooperators is minimized when $\pi_1 = 0$. Conversely, for $p_A\in [p_A^*, 1)$, $G_2(p_A)\leq0$ always satisfies, and thereby the fitness difference between defectors and cooperators is minimized when $\pi_1 = 1$.
$\hfill\blacksquare$

\begin{center}
\begin{figure*}[t]
\centering
\includegraphics[scale = 0.7]{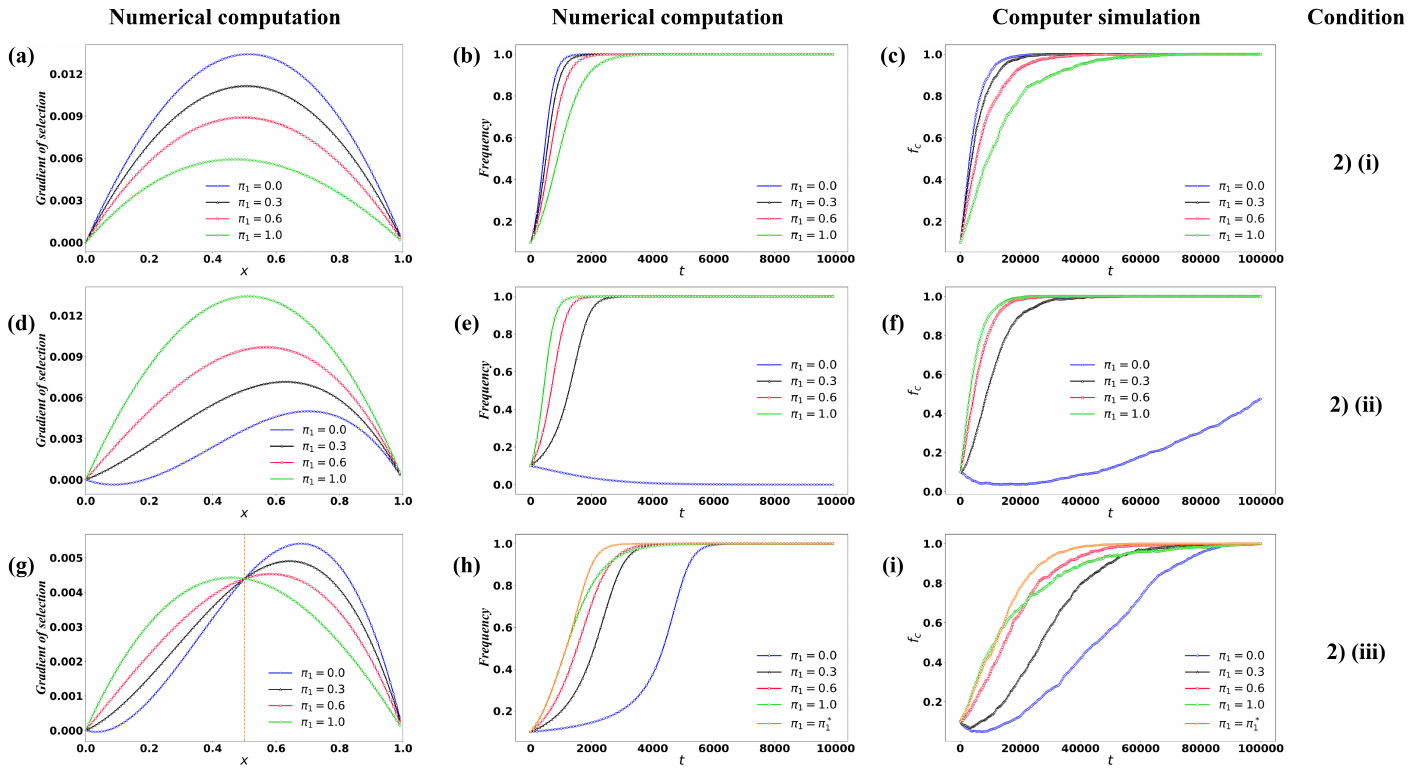}
\caption{\textbf{Numerical and simulation results for maximizing the gradient of cooperation selection.} The first column illustrates the numerical results of the gradient of cooperation selection as a function of the fraction of cooperators under different stationary game distributions. The second and third columns display the corresponding numerical and Monte Carlo simulation results, respectively, for the evolutionary trajectories of the cooperator frequency over time. Each row corresponds to a different theoretical case for maximizing the gradient of cooperation selection, as stated in items 2)(i)-(iii) of Thm.~3 in the main manuscript. The vertical orange dashed line in panel~(g) denotes the critical point $x^*$ at which $G_1(x^*) = 0$ shown in Eq.~(33).}
\label{gradient-supp}
\end{figure*}
\end{center}

\section{Further Numerical and Simulation Results for Optimization}

In this section, we present additional numerical and simulation results about the optimization problems of (i) maximizing the gradient of cooperation selection and (ii) minimizing the fitness difference between defectors and cooperators. These results serve to validate the final three theoretical predictions stated in Thms.~3 and 4 of the main manuscript, respectively.

\subsection{Further Numerical and Simulation Results for Maximizing the Gradient of Cooperation Selection}

To validate the final three theoretical predictions regarding the maximization of the gradient of cooperation selection as demonstrated in Thm.~3 of the main manuscript, we provide corresponding numerical and simulation results for each of the three cases in Fig.~\ref{gradient-supp}. Specifically, the parameter settings for rows 1-3 in the figure satisfy $Dg_2 + Dr_1 < Dg_1 + Dr_2$, with $(k^2-k-1)(Dr_1-Dr_2)+(Dg_1-Dg_2)>0$ for row~1, $(k^2-k-1)(Dg_1-Dg_2)+(Dr_1-Dr_2)<0$ for row~2, and $(k^2-k-1)(Dr_1-Dr_2)+(Dg_1-Dg_2)<0, (k^2-k-1)(Dg_1-Dg_2)+(Dr_1-Dr_2)>0$ for row~3.

 In Fig.~\ref{gradient-supp}(a), we observe that the gradient of selection corresponding to $\pi_1 = 0.0$ marked by blue circles is the highest across all values of cooperator frequency and decreases monotonically as $\pi_1$ increases. Fig.~\ref{gradient-supp}(b) shows that all scenarios ultimately lead to full cooperation, with $\pi_1 = 0.0$ achieving this most rapidly. As $\pi_1$ increases, the time required to reach full cooperation increases. These trends are further confirmed by the Monte Carlo simulation results in Fig.~\ref{gradient-supp}(c). In contrast, Fig.~\ref{gradient-supp}(d) demonstrates that the gradient of selection is maximized for $\pi_1 = 1.0$ marked by the green diamonds and decreases as $\pi_1$ decreases. This is corroborated by Fig.~\ref{gradient-supp}(e), which shows that cooperation emerges most quickly under $\pi_1 = 1.0$, whereas in the $\pi_1 = 0.0$ scenario, cooperators will eventually go extinct. These observations are again supported by the simulation results illustrated in Fig.~\ref{gradient-supp}(f). In Fig.~\ref{gradient-supp}(g), a threshold phenomenon is observed: when the cooperator frequency $x \in (0, x^*]$, where $x^*=[( k^2-k-1 ) ( Dr_1-Dr_2 ) +( Dg_1-Dg_2 )] / [( k^2-k-2 ) ( Dr_1-Dg_1-Dr_2+Dg_2 )]$, indicated by the orange vertical dashed line, the gradient of selection is largest at $\pi_1 = 1.0$; however, when $x \in (x^*, 1)$, it is maximized at $\pi_1 = 0.0$. Furthermore, both the numerical and simulation results presented in Figs.~\ref{gradient-supp}(h) and \ref{gradient-supp}(i) demonstrate that, given a sufficiently long evolutionary time, the cooperator frequency under $\pi_1 = \pi_1^*$ always achieves the highest or reaches the full cooperation state fastest compared to the other cases, where $\pi_1^*$ denotes the value of $\pi_1$ that maximizes the gradient of selection in Fig.~\ref{gradient-supp}(g). These results provide strong support for the theoretical predictions, with numerical and simulation outcomes in excellent agreement.

\subsection{Further Numerical and Simulation Results for Minimizing the Fitness Difference between Defectors and Cooperators}

\begin{center}
\begin{figure*}[htbp]
\centering
\includegraphics[scale = 0.7]{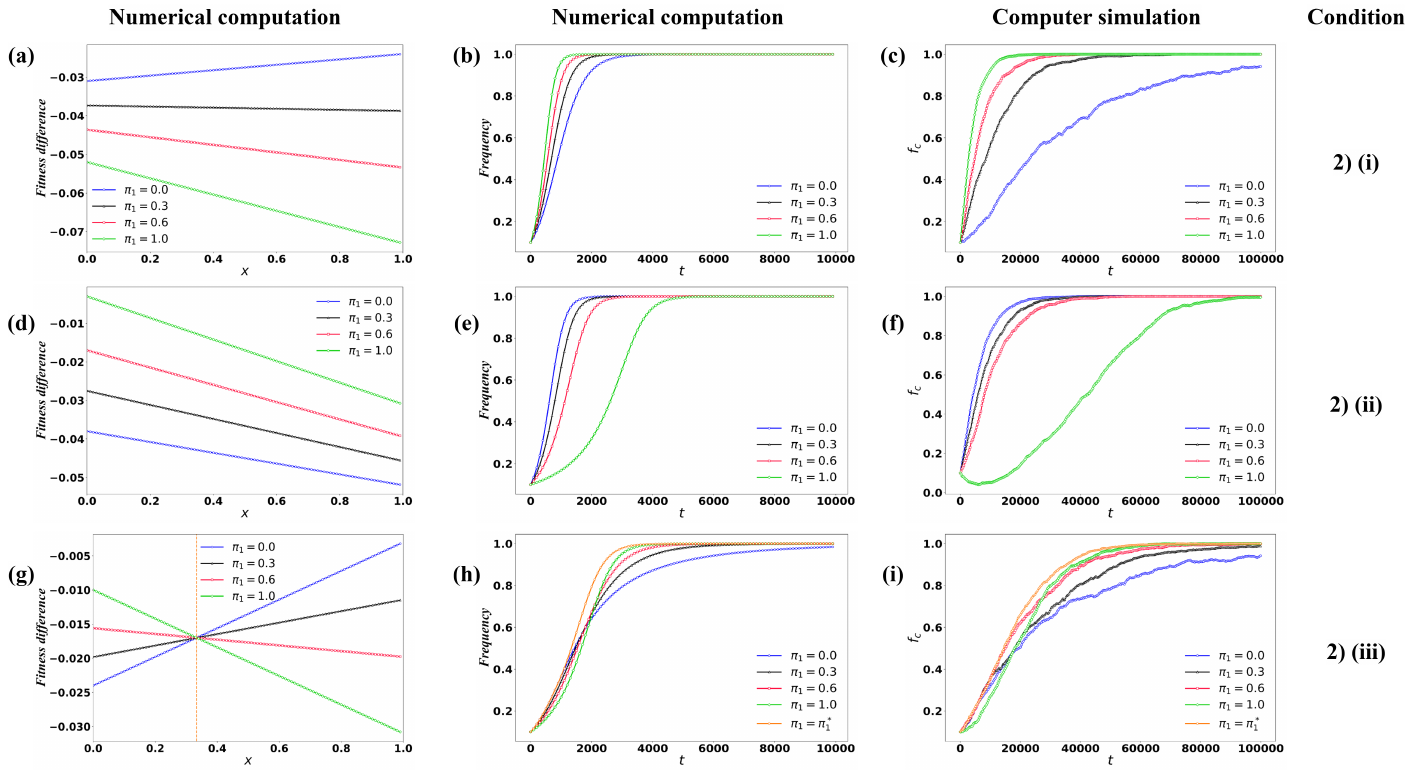}
\caption{\textbf{Numerical and simulation results for minimizing the fitness difference between defectors and cooperators.} The first column represents the numerical results of the fitness difference between defectors and cooperators as a function of the proportion of cooperators under different stationary game distributions. The second and third columns display the numerical and simulation results, respectively, for the evolutionary trajectories of the cooperator frequency over time under these distributions. Each row corresponds to a different theoretical case for minimizing the fitness difference between defectors and cooperators, as described in items 2)(i)-(iii) of Thm.~4 in the main manuscript. The vertical orange dashed line in panel~(g) indicates the critical point $x^*$ that makes $G_2(x^*) = 0$ presented in Eq.~(37).}
\label{fitness difference-supp}
\end{figure*}
\end{center}

To validate the final three theoretical predictions regarding the minimization of the fitness difference between defectors and cooperators as presented by Thm.~4 of the main manuscript, we provide corresponding numerical and simulation results for each of the three cases in Fig.~\ref{fitness difference-supp}. Concretely, the parameter settings satisfy $Dg_1 + Dr_2 < Dg_2 + Dr_1$, with $Dr_1 < Dr_2$ for row~1, $Dg_1 > Dg_2$ for row~2, and $Dr_1 > Dr_2, Dg_1 < Dg_2$ for row~3.

In Fig.~\ref{fitness difference-supp}(a), we observe that the fitness difference between defectors and cooperators is minimized at $\pi_1 = 1.0$ marked by green diamonds and increases monotonically as $\pi_1$ decreases. Fig.~\ref{fitness difference-supp}(b) further shows that all scenarios eventually reach the fully cooperative state, with the fastest convergence occurring at $\pi_1 = 1.0$. The time required to achieve full cooperation increases as $\pi_1$ decreases. These findings are corroborated by the Monte Carlo simulation results displayed in Fig.~\ref{fitness difference-supp}(c). Conversely, Fig.~\ref{fitness difference-supp}(d) shows that the fitness difference is minimized at $\pi_1 = 0.0$ indicated by blue circles and increases with increasing $\pi_1$. Analogously, Fig.~\ref{fitness difference-supp}(e) demonstrates that the fully cooperative state is reached most rapidly when $\pi_1 = 0.0$, a conclusion also supported by the simulation results shown in Fig.~\ref{fitness difference-supp}(f). In Fig.~\ref{fitness difference-supp}(g), we identify a critical threshold $x^*=\left( Dr_2-Dr_1 \right) / \left( Dg_1-Dg_2-Dr_1+Dr_2 \right)$ marked by the orange vertical dashed line. When the cooperator frequency $x \in (0, x^*]$, the fitness difference is smallest for $\pi_1 = 0.0$; in contrast, for $x \in (x^*, 1)$, the smallest fitness difference occurs at $\pi_1 = 1.0$. Furthermore, the numerical and simulation results depicted in Figs.~\ref{fitness difference-supp}(h) and \ref{fitness difference-supp}(i) reveal that over a sufficiently long evolutionary period, the scenario with $\pi_1 = \pi_1^*$ achieves the highest number of cooperators or reaches full cooperation the fastest compared to the other cases, where $\pi_1^*$ corresponds to the value of $\pi_1$ that minimizes the fitness difference in Fig.~\ref{fitness difference-supp}(g). Collectively, these results exhibit strong consistency between theoretical predictions and both numerical and simulation findings.

\ifCLASSOPTIONcaptionsoff
  \newpage
\fi